**Isotope effect in the Morse approximation of the ground state term of hydrogen molecule $^nH_2$, $n$ = 1÷7. Herzberg anomaly and anharmonicity**

Gleb S. Denisov[a]; Edem R. Chakalov[b]; Peter M. Tolstoy[b*]

[a]Department of Physics, St. Petersburg State University, Russia

[b]Institute of Chemistry, St. Petersburg State University, Russia

[*]E-mail: peter.tolstoy@spbu.ru

## Abstract

The influence of the reduced mass on the results of the approximation by the Morse function of the ground electronic state term potential U($r$) of hydrogen molecule $^nH_2$ for seven symmetrical isotopologues $n$ = 1–7 is investigated. For each isotopologue, two alternative model solutions of the Schrödinger equation are obtained, namely, M1($r$) and M2($r$) functions, which differ by the selection of the primary fitting parameters in the Morse approximation. The U($r$) – M($r$) functions are shown to clearly visualize the differences between the shapes of the potential at its approximation. The concept of systematic deviation of Morse functions from the real potential is introduced. This deviation is always present in the approximation of simple terms, for which U($r$) curve lies between M1($r$) and M2($r$) without crossings. In case of $^nH_2$, for M1($r$) the systematic monotonous deviation increases, reaching 1700–2770 cm$^{-1}$ (for $^7H_2$–$^1H_2$) at the level of the asymptote. For M2($r$) the systematic deviation decreases at larger distances: it is dome-shaped with a maximum of ca. 1500 cm$^{-1}$ in the upper part of the term, after which it falls to zero. The M2($r$) curves for different isotopologues diverge by less than 20 cm$^{-1}$. There are also anomalies, where M1($r$) and M2($r$) cross U($r$), called "Herzberg anomaly". For example, at around $r$ = 1.08 Å the U($r$) – M2($r$) function shows anomalous negative values ranging from –125 cm$^{-1}$ (for $^1H_2$) to –142 cm$^{-1}$ (for $^7H_2$). The diagnostic value of Herzberg anomaly and anharmonicities for various isotopologues is discussed.

## 1. Introduction

In papers [1–5], the features of two alternative approximations of the potential function of a diatomic molecule by the Morse function M($r$) are considered. The reason for this ambiguity lies in the impossibility to use a single parameter for the rigorous introduction of two new quantities, which is required for the simplest transition to the anharmonic description of the vibrational motion. These two new quantities are the binding energy $D_e$, which determines the limiting value of the energy of bound states, and the anharmonicity of vibrations $x$*, which determines the width of the potential well. In the harmonic description, the frequency of vibration simultaneously gives the energy of the vibrational quantum, which depends on the width of the potential well. Morse managed to find a form of potential, for which the solution of the Schrödinger equation gives in a good approximation the energy eigenvalues in the form of a binomial described by constants $\omega_e$ and $\omega_e x_e$. These quantities are related to the energy $D_e$ by the approximate equation

$$D_e = \omega_e{}^2/4\omega_e x_e \qquad (1)$$

which is valid only when the expression for energy is strictly two-term, that is, when there are no cubic and subsequent terms in the series. Since there are no molecules in nature for which the convergence of vibrational levels is described by a single parameter, using equation (1) requires to choose between two possibilities. Morse in [6] gives a variant in which the function M($r$) is constructed from a known value of anharmonicity; in [1] this function was called M1($r$). In this case, the binding energy $D_e$ as defined in (1) is subject to large errors, because it is extrapolated over a large distance; the accuracy with which M1($r$) approximates the shape of the real potential function – denoted U($r$) throughout this paper – worsens monotonously from the bottom of the well to the asymptote. If the $D_e$ value is known it could be used to construct M2($r$) (this option is not mentioned in Morse's paper, but is easily implied). The accuracy of the M2($r$) approximation is higher at the bottom and at the asymptote of the potential, so that the systematic error is dome-shaped, reaching maximum somewhere in the upper half of U($r$).[5] To construct either of Morse functions, M1($r$) or M2($r$), it is also necessary to know the equilibrium distance $r_e$ and the reduced mass $\mu$. Equation (1) is used also to construct alternative values of the Morse parameter "$a$", see below, Section 3.1.

The nature of the distortion of the shape of U($r$) when approximated by the Morse function M2($r$) was shown by Herzberg on the example of the ground electronic term X $^1\Sigma_g{}^+$ of the hydrogen molecule $^1H_2$ in Fig. 48 in [7] (see also the dashed line in Fig. S1 in Supporting information). In the lower part of U($r$) its width was shown to be increased comparing to the fitted Morse function up until their intersection at 1.5 Å. In [1] this was called the Herzberg anomaly, and it is analyzed in detail in [3]. It turns out to be several times smaller in magnitude than shown originally by Herzberg in his book. For the second time after Herzberg, the result of Morse analysis of the $^1H_2$ potential curve constructed by the pseudospectral method was shown in Fig. 4 in [8]. The result was similar to Herzberg's result, but the intersection of M2($r$) with U($r$) occurred at noticeably larger interatomic distances, $r$ ~ 1.8 Å. The Herzberg anomaly was also found computationally in the analysis of ground electronic terms of several other molecules, such as $O_2$ and HCl,[4,5] which makes one wonder about the nature of the anomaly.

The present work has been undertaken to approximate the U($r$) potential for the ground electronic term X $^1\Sigma_g{}^+$ of $^nH_2$ by the Morse functions M1($r$) and M2($r$) for all known hydrogen isotopes $n$ = 1–7, and to find out the changes occurring in the shapes of the resulting curves and vibrational level systems. Note that the lifetime of isotopes $^4H_2$–$^7H_2$ is too short for the formation of a molecule with defined structure of vibrational states, so that these molecules were treated hypothetically in order to check which tendencies in the Morse approximations of the sequence $^1H_2 \rightarrow {}^2H_2 \rightarrow {}^3H_2$ are persistent and which one are accidental. For instance, the dense structure of vibrational energy states for heavier isotopes should allow one to probe finer details of the U(*r*) potential. In other words, the actual existence of the isotopes studied is not important, since we studied solely the effect of the reduced mass on the M1(*r*) and M2(*r*) models. The $H_2$ molecule is considered in the Born-Oppenheimer approximation, using the theoretically calculated potential curve U(*r*), which does not depend on the mass of the isotope. Special attention is paid to the behavior of Herzberg anomaly for the series of isotopologues.

## 2. Methods

As the potential U(*r*) for further analysis and approximations with Morse functions we have chosen one of the most accurate $H_2$ potentials available in literature,[9–12] which was obtained by Pachucki[12] for $^1H_2$ (in this work, 2048 points with ~0.005 Å increment are obtained from the potential given in [12] by polynomial interpolation within the 0.265–10.584 Å range of interatomic distances, see Table S1 in Supporting information). This potential U(*r*) is used throughout the paper and it is the same for all the $^nH_2$ isotopologues ($n$ = 1–7) assuming perfect fulfillment of Born-Oppenheimer approximation. U(*r*) is characterized by the bond energy $D_e$ = 38293.017 cm$^{-1}$ and the equilibrium interatomic distance $r_e$ = 0.74144 Å. In Table S2, accurate theoretical values of $r_e$ for $^1H_2$ and vibrationally averaged $\langle r \rangle$ interatomic distances for $^nH_2$, $n$ = 1–3, available in literature, are given for comparison. The overall trend is a slight decrease of $\langle r \rangle$ upon increase of the mass of $n$.

For $^nH_2$ isotopologues, $n$ = 1–7, the Schmidt's WavePacket MATLAB package[13,14] was employed for numerical solution of the vibrational time-independent Schrödinger equation with the potential U(*r*) by diagonalization of the Hamiltonian 2048×2048 matrix based on the discrete variable representations.[15,16] The resulting values of vibrational energy levels G($\upsilon$) for all isotopologues are given in Table S3.

In Table 1, for each of the $^nH_2$ isotopologues, the reduced masses $\mu$, the quantum numbers for the highest vibrational levels $\upsilon_{max}$, the dissociation energies from the ground vibrational level $D_0$, the positions of the turning points for the $\upsilon_{max}$ levels, and the energy differences between the $\upsilon_{max}$ level and the dissociation limit $D_e$ are given.

Table 1. Reduced masses $\mu$, calculated dissociation energies $D_0$ from the zero-point vibrational levels, turning points for the levels with the highest vibrational quantum numbers $\upsilon_{max}$ and energy differences between the asymptote $D_e$ and the $\upsilon_{max}$ levels for $^1H_2$–$^7H_2$ in the U($r$) potential.

| | Reduced mass $\mu$, a.u. | $D_e - G(0) \equiv D_0$, cm$^{-1}$ | $\upsilon_{max}$ | Turning points for $\upsilon_{max}$, Å | $D_e - G(\upsilon_{max})$, cm$^{-1}$ |
|---|---|---|---|---|---|
| $^1H_2$ | 918.571 | 36112.113 | 14 | 0.41582, 3.26903 | 140.660 |
| $^2H_2$ | 1835.730 | 36745.517 | 21 | 0.41078, 5.57782 | 1.293 |
| $^3H_2$ | 2748.944 | 37026.883 | 25 | 0.41078, 3.55133 | 64.502 |
| $^4H_2$ | 3669.844 | 37196.318 | 30 | 0.41078, 6.59106 | 0.416 |
| $^5H_2$ | 4589.376 | 37311.761 | 33 | 0.41078, 4.06047 | 17.973 |
| $^6H_2$ | 5509.610 | 37397.137 | 36 | 0.41078, 3.84371 | 30.264 |
| $^7H_2$ | 6428.149 | 37463.199 | 39 | 0.41078, 3.90924 | 25.655 |

## 3. Results and discussion

### 3.1. Approximations of the calculated potential

We have constructed an approximation of the potential function U($r$) by the Morse function

$$\mathrm{M}(r) = D_\mathrm{e}\left[1 - e^{-a(r-r_\mathrm{e})}\right]^2 \qquad (2)$$

in two ways, M1($r$) and M2($r$), which differ in the definition of the $D_e$ and $a$ parameters. M1($r$) was defined using $a = (2\mu\omega_\mathrm{e}x_\mathrm{e})^{\frac{1}{2}}$ and $D_\mathrm{e} = \omega_\mathrm{e}{}^2/(4\omega_\mathrm{e}x_\mathrm{e})$. M2($r$) was defined using $a = \omega_\mathrm{e}(\mu/2D_\mathrm{e})^{\frac{1}{2}}$ and the exact value of the binding energy for U($r$). In both cases, $\mu$ is the reduced mass of a particular isotopologue $^nH_2$. The harmonic frequencies $\omega_e$ and anharmonicities $\omega_e x_e$ were calculated according to Eq. (3):

$$G(\upsilon) = \omega_\mathrm{e}\left(\upsilon + \frac{1}{2}\right) - \omega_\mathrm{e}x_\mathrm{e}\left(\upsilon + \frac{1}{2}\right)^2 \qquad (3)$$

using the numerical values of the $0 \rightarrow 1$ and $0 \rightarrow 2$ transition frequencies collected in Table S3. The main difference between the M1($r$) approximations for various isotopologues is near the asymptote and in the region of the continuous spectrum. For M2($r$) approximations the main difference is at the top region of the attractive branch of U($r$), while the asymptotes of M2($r$) coincide with those of U($r$) by definition. Numerical parameters of the M1($r$) and M2($r$) models for all $^nH_2$ isotopologues are collected in Tables 2 and 3, respectively. The G($\upsilon$) values for M1($r$) and M2($r$) models are given in Tables S4 and S5, respectively, for all $^nH_2$ isotopologues.

Table 2. Approximation parameters ($\omega_e$, $\omega_e x_e$, $x^*$ and $a$) of the M1($r$) potential for $^1H_2$–$^7H_2$ with the corresponding dissociation energies $D_0$ from the zero-point vibrational levels, turning points for the levels with the highest vibrational quantum numbers $\upsilon_{max}$ and energy differences between the asymptote $D_e$ and the $\upsilon_{max}$ levels.

| | $\omega_e$, cm$^{-1}$ | $\omega_e x_e$, cm$^{-1}$ | $x^*$ | $a$, Å$^{-1}$ | $D_e$, cm$^{-1}$ | $D_0 \equiv D_e - G(0)$, cm$^{-1}$ | $\upsilon_{max}$ | Turning points for $\upsilon_{max}$, Å | $D_e - G(\upsilon_{max})$, cm$^{-1}$ |
|---|---|---|---|---|---|---|---|---|---|
| $^1H_2$ | 4400.0274 | 117.8579 | 0.02679 | 1.87697 | 41066.922 | 38896.373 | 18 | 0.37549, 6.13737 | 3.274 |
| $^2H_2$ | 3114.5645 | 59.5874 | 0.01913 | 1.88670 | 40698.697 | 39156.311 | 26 | 0.37549, 5.63327 | 7.963 |
| $^3H_2$ | 2546.1252 | 40.0541 | 0.01573 | 1.89290 | 40462.464 | 39199.415 | 31 | 0.37549, 6.09200 | 3.221 |
| $^4H_2$ | 2203.7448 | 30.0680 | 0.01364 | 1.89495 | 40379.199 | 39284.844 | 36 | 0.38053, 6.93889 | 0.641 |
| $^5H_2$ | 1970.6627 | 24.0325 | 0.01220 | 1.89451 | 40398.586 | 39419.262 | 40 | 0.38053, 5.75929 | 6.008 |
| $^6H_2$ | 1798.8141 | 20.0819 | 0.01116 | 1.89751 | 40281.640 | 39387.253 | 44 | 0.38053, 6.42975 | 1.653 |
| $^7H_2$ | 1665.5930 | 17.3385 | 0.01041 | 1.90445 | 40000.586 | 39172.124 | 48 | 0.38053, 5.96597 | 3.803 |

Table 3. Approximation parameters ($\omega_e$, $\omega_e x_e$, $x^*$ and $a$) of the M2($r$) potential for $^1H_2$–$^7H_2$ with the corresponding dissociation energies $D_0$ from the zero-point vibrational levels, turning points for the levels with the highest vibrational quantum numbers $\upsilon_{max}$ and energy differences between the asymptote $D_e$ and the $\upsilon_{max}$ levels.

| | $\omega_e$, cm$^{-1}$ | $\omega_e x_e$, cm$^{-1}$ | $x^*$ | $a$, Å$^{-1}$ | $D_e$, cm$^{-1}$ | $D_0 \equiv D_e - G(0)$, cm$^{-1}$ | $\upsilon_{max}$ | Turning points for $\upsilon_{max}$, Å | $D_e - G(\upsilon_{max})$, cm$^{-1}$ |
|---|---|---|---|---|---|---|---|---|---|
| $^1H_2$ | 4400.0274 | 126.3954 | 0.02873 | 1.94376 | | 36124.602 | 17 | 0.38557, 6.46503 | 1.121 |
| $^2H_2$ | 3114.5645 | 63.3308 | 0.02033 | 1.94506 | | 36751.568 | 24 | 0.38557, 6.86831 | 0.509 |
| $^3H_2$ | 2546.1252 | 42.3233 | 0.01662 | 1.94578 | | 37030.536 | 30 | 0.38557, 5.48204 | 7.486 |
| $^4H_2$ | 2203.7448 | 31.7061 | 0.01439 | 1.94588 | 38293.017 | 37199.071 | 34 | 0.38557, 6.15753 | 2.024 |
| $^5H_2$ | 1970.6627 | 25.3539 | 0.01287 | 1.94590 | | 37314.024 | 38 | 0.38557, 5.90044 | 3.342 |
| $^6H_2$ | 1798.8141 | 21.1248 | 0.01174 | 1.94616 | | 37398.891 | 42 | 0.38557, 7.59926 | 0.122 |
| $^7H_2$ | 1665.5930 | 18.1117 | 0.01087 | 1.94645 | | 37464.749 | 45 | 0.38557, 5.77946 | 4.195 |

Fig. 1a shows the shape of U($r$) and its M1($r$) models constructed for all isotopologues $n$ = 1–7. Due to the long-range extrapolation, the value of the binding energy $D_e$ for M1($r$) approximations is dramatically overestimated in comparison with the actual value for U($r$). A maximum deviation of 2774 cm$^{-1}$ is obtained for the lightest $^1H_2$ isotope, while the entire range of asymptotes for $n$ = 1–7 is ca. 1070 cm$^{-1}$ wide. The difference of ca. 1700 cm$^{-1}$ constitutes the systematic deviation (systematic error) of M1($r$). It is noteworthy how the positions of the asymptotes change: going from $n$ = 1 to $n$ = 2, $D_e$ decreases by ca. 370 cm$^{-1}$, and for heavier isotopes the changes become smaller and irregular (see inset I in Fig. 1a). The difference in the asymptote positions is due to the fact that M1($r$) was constructed using the positions of the first three vibrational levels, which are located in different parts of the potential curve for different isotopologues, namely, the closer to the minimum the heavier the isotopologue. Besides the monotonic decrease of the energy $D_e$ with increasing $n$, the zero-point vibrational level also decreases. As a result, the dissociation energy $D_0$ increases with increasing mass (see Table 2). Previously, a similar trend was observed in [17,18] when comparing the first three members of the series, namely $^1H_2$, $^2H_2$ and $^3H_2$. In the whole $^1H_2$–$^7H_2$ series, the $D_0$ increase is 276 cm$^{-1}$. Insets I and II of Fig. 1a indicate that M1($r$) curves for $n$ = 4 and $n$ = 5 are interchanged, which is confirmed by the analysis of the data in Table 2: the monotonic sequence of changes is broken for these isotopologues. Within the series, the values of the reduced dimensionless anharmonicity $x^*$ monotonously decrease by more than half, from 0.029 to 0.011. Such a dramatic change deserves particular consideration. So far, we have not encountered any literature data on the behavior of anharmonicity $x^*$. We assume that the parameters of the M1($r$) functions for $n$ = 4 and $n$ = 5 are very close due to the small change in the anharmonicity of the initial potential U($r$) in the region where the first three vibrational levels for these two isotopologues are located. Thus, we believe that the mathematically detectable non-monotony of the M1($r$) parameters when shifting from $n$ = 4 to $n$ = 5 has no special physical meaning. According to Table 2, the number of excess vibrational levels for M1($r$) potentials in the $n$ = 1–7 series increases by 2–12, and within the limits of the U($r$) potential well this number decreases by 1–3.

Fig. 1b shows the difference functions $\delta 1(r)$ = U($r$) – M1($r$) for $^1H_2$–$^7H_2$. The extrema of the $\delta 1(r)$ curves indicate significant deviations of the potential U($r$) from its M1($r$) models, since a monotonous increase of the systematic error is usually observed for molecules characterized by simple electronic terms.[5] The slowly diverging bundle of curves for $n$ = 1–7 follows the attractive branch of U($r$) in a wavy manner, forming extrema at ca. 1.25 Å (inset I in Fig. 1b) and ca. 1.9 Å (inset II). For $n$ = 3–7 the $\delta 1(r)$ curves at around 1.9 Å become positive. Consequently, there are two intersections between U($r$) and M1($r$), so that a part of the M1($r$) curve appears on the outer side of the attraction branch of U($r$). For $^7H_2$ isotopologue, the width of this region is largest, spanning from 1.529 Å to 2.370 Å. Inset II in Fig. 1a shows the arrangement of M1($r$) curves in the region of such an anomalous behavior.

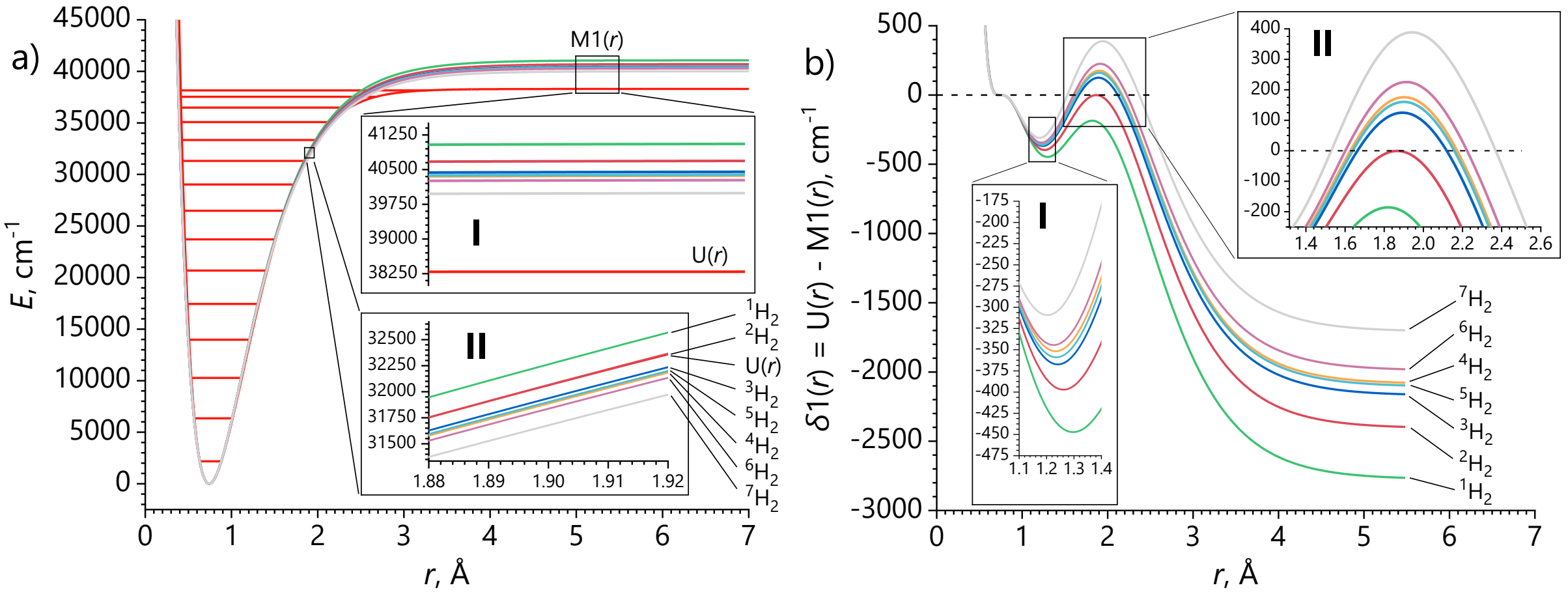

Figure 1. Comparison of the calculated potential U($r$) with its M1($r$) approximations for $^1H_2$–$^7H_2$: (a) U($r$) with a set of its own vibrational levels (obtained as described in Methods) and a sequence of M1($r$) functions and (b) $\delta 1(r)$ = U($r$) – M1($r$). Insets in (a) show the deviation of M1($r$) from U($r$) for two selected regions; insets in (b) show the regions of extrema of $\delta 1(r)$.

Fig. 2a shows the shapes of the potential curve U($r$) and its M2($r$) models constructed for $^1H_2$–$^7H_2$. A systematic dome-shaped deviation is observed for $r$ = 1.3–4.0 Å before M2($r$) functions merge with the common asymptote. The magnitude of the deviation reaches its maximum about 1500 cm$^{-1}$ at around 2.09 Å. The inset in Fig. 2a indicates that the deviation monotonously decreases when going from $n$ = 1 to $n$ = 7. Unlike M1($r$), the M2($r$) curves for $n$ = 4 and $n$ = 5 are not interchanged, although the difference between them is less than 1 cm$^{-1}$.

Fig. 2b shows the difference functions $\delta2(r)$ = U($r$) – M2($r$). In the region 0.75–1.25 Å (inset II) a deviation from U($r$) with a depth up to 142 cm$^{-1}$ is observed; this is the Herzberg anomaly. This anomaly affects the position of vibrational levels, which was observed spectroscopically as early as 1959 by Herzberg and Howe,[19] who urged theorists to address this phenomenon. After crossing zero, $\delta2(r)$ curves manifest a dome-shaped systematic deviation, which falls to zero when U($r$) and M2($r$) reach the asymptote. For the series of isotopologues, the isotope effect is small: the differences in the heights of extrema at 1.09 Å and 2.09 Å are similar, ca. 18 cm$^{-1}$ (compare insets I and II) and $\delta2(r)$ curves move to lower energy by a parallel shift as a whole with increasing $n$.

All intersection points of the U($r$) potential with its M1($r$) and M2($r$) approximations for $^{1}H_2$–$^{7}H_2$ are collected in Table S6.

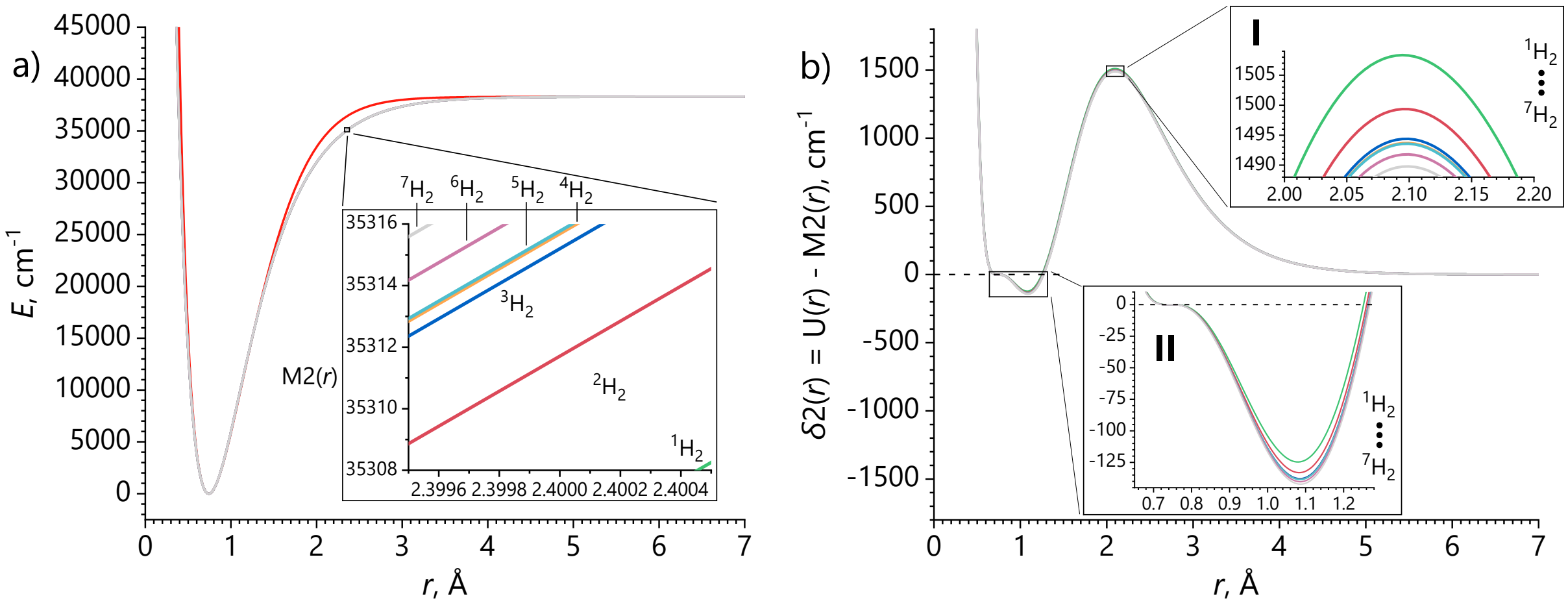

Figure 2. Comparison of the calculated potential U($r$) with its M2($r$) approximation for $^{1}H_2$–$^{7}H_2$: (a) U($r$), M2($r$) and (b) U($r$) – M2($r$).

The last two columns of Tables 2 and 3 provide positions of the turning points for the upper vibrational levels $\upsilon_{max}$, as well as the energy differences between the $\upsilon_{max}$ levels and the binding energy $D_e$. Predictably, the system of vibrational levels for M2($r$) model, which accurately reproduces the $D_e$ energy for the U($r$) potential, describes this potential at large internuclear distances much better than the system of vibrational levels for M1($r$) model. The outer turning points for M1($r$) lie in the region up to 7.0 Å, while for M2($r$) they lie in the region up to 7.6 Å. However, the dependence on the isotopologue mass is not monotonous: the turning point appears at large distances when the upper vibrational level "accidentally" appears extremely close to the dissociation limit, which is observed only for $n$ = 2 and $n$ = 6.

## 3.2. Isotope effect on anharmonicity

Accurate calculated data on the vibrational structure of $^1H_2$, $^2H_2$, and $^3H_2$ isotopologues are given by Wolniewicz,[17,20] Pachucki,[18,21,22] Bubin,[23–26] Stanke,[27,28] Komasa[29,30] *et al*. for 15, 21 and 26 energy levels, respectively. Paper [2] provides analysis of the isotope effect on the anharmonicity of $^nH_2$ for $n$ = 1–3 based on data from [17], and for the HF molecule based on data from [31]. Following up on the important work of McCoy,[32] the role of abscissa in anharmonicity analysis was investigated in [2], and it was shown that the energy scale as a characteristic of the potential curve allows one to more adequately compare the data for various isotopologues than the vibrational quantum number scale $\upsilon$. As a continuation of the abovementioned studies, in the present work, Figs. 3a,b show the values of the anharmonicity function $-2\omega_e x(\upsilon)$ *vs.* $\upsilon$ and $E \equiv G(\upsilon) - D_e$ coordinates for seven isotopologues $^nH_2$ in the U($r$) potential. Here, again, the choice of energy as a unified scale for all isotopologues is physically more reasonable, since it demonstrates the distribution of $-2\omega_e x(\upsilon)$ points in a constant energy range $D_e$. The anharmonicities $-2\omega_e x(\upsilon)$ for each $\upsilon$ were calculated by equation (3) using three successive vibrational levels with quantum numbers $\upsilon - 1$, $\upsilon$ and $\upsilon + 1$.

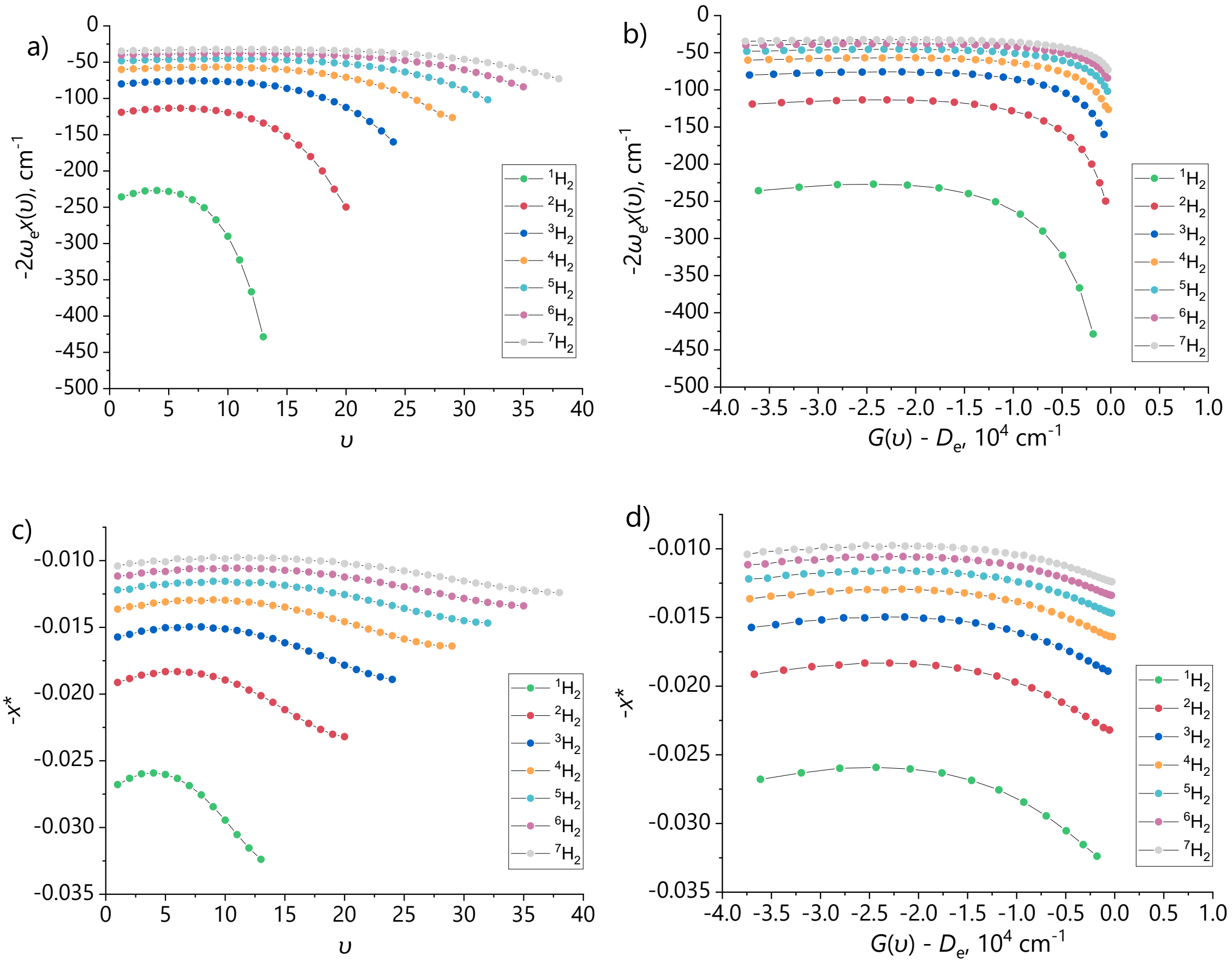

Figure 3. (a,b) Anharmonicities $-2\omega_e x(\upsilon)$ and (c,d) reduced anharmonicities $-x^*$ *vs.* vibrational quantum number $\upsilon$ (a,c) and vibrational energies $E \equiv G(\upsilon) - D_e$ (b,d) of $^1H_2$–$^7H_2$.

It can be seen that the anharmonicity $-2\omega_e x(\upsilon)$ has a non-monotonous character, and in the lower part of the potential reveals an anomalous region of decrease in absolute value. Meanwhile, the decrease of anharmonicity $-2\omega_e x(\upsilon)$ with increasing $n$ is caused by the shrinking of the region on the potential curve where the three corresponding levels are located. The non-monotonous nature of anharmonicity was noted in [19] when analyzing the emission spectra of $^1H_2$, though instead of "anharmonicity" the authors used the equivalent term "second difference", $\Delta_2G(\upsilon) = \Delta_1G(\upsilon) - \Delta_1G(\upsilon - 1)$, where $\Delta_1G(\upsilon) = G(\upsilon + 1) - G(\upsilon)$. In [2], we assumed that the non-monotony of $-2\omega_e x(\upsilon)$ is related to the Herzberg anomaly and undertook the analysis of isotope effects in the $\Delta_2G(\upsilon)$ dependences for $^2H_2$ and $^3H_2$ using data from [17]. In the present work, these dependences are extended to the complete set of isotopologues from $^1H_2$ to $^7H_2$. Fig. 3b indicates that with the increase of $n$, the convexity of the curves decreases, and the concept of anomaly seems to lose its meaning. However, this is due to the fact that both the transition frequencies $\Delta_1G(\upsilon)$ and the second differences $\Delta_2G(\upsilon)$ decrease in absolute value as the levels approach each other for the heavier isotopologues. If the anharmonicity values are multiplied by the reduced mass ratio $\mu(^nH_2)/\mu(^1H_2)$,[2] they coincide with good accuracy, see Fig. 4.

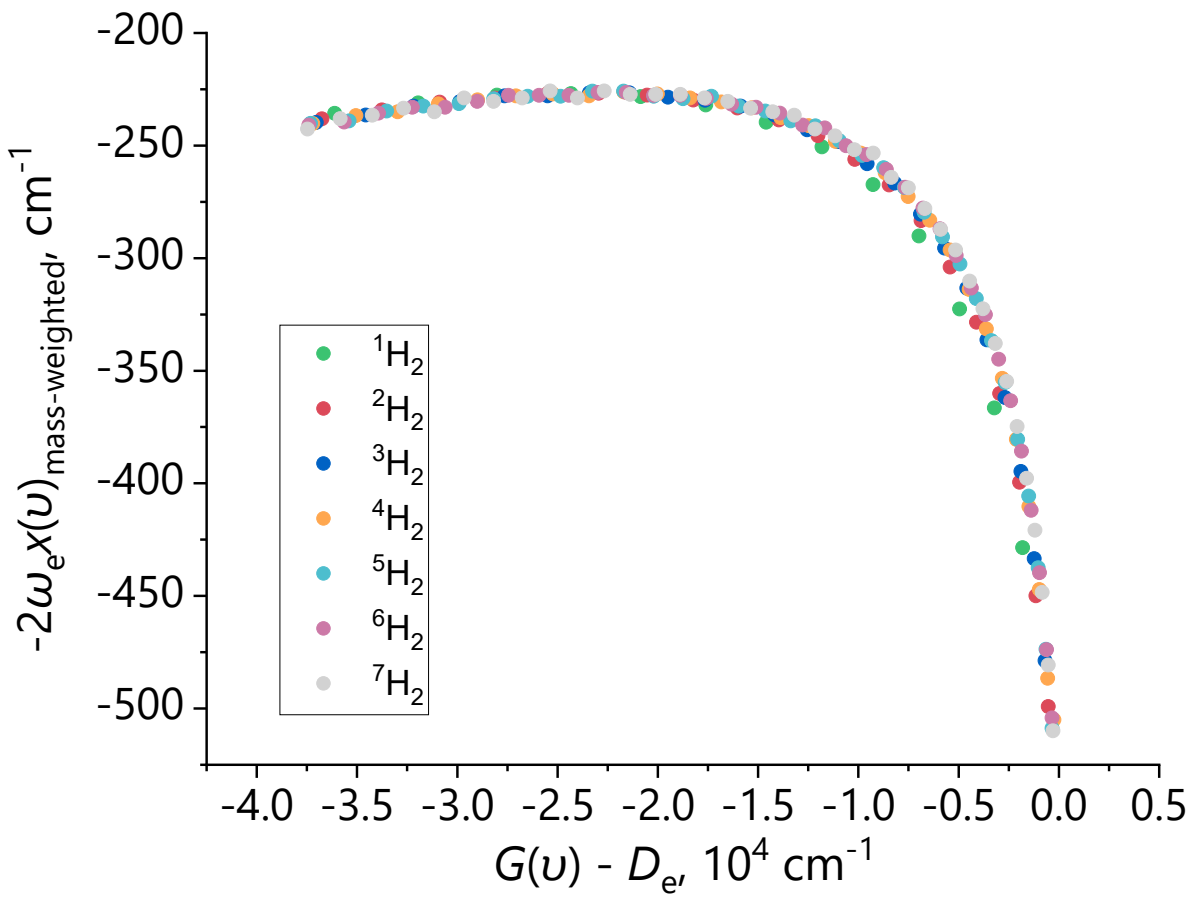

Figure 4. Mass-weighted anharmonicities $-2\omega_e x(\upsilon)_{mass\text{-}weighted} = -2\omega_e x(\upsilon)\cdot\mu(^nH_2)/\mu(^1H_2)$ *vs.* vibrational energies $E \equiv G(\upsilon) - D_e$ of $^1H_2$–$^7H_2$.

Also in [2], an attempt was made to construct a universal (dimensionless) frequency-independent anharmonicity function $x^*(\upsilon)$ (or $x^*(E)$) for $n$ = 1–3. Similar functions are plotted in Figs. 3c,d for $^nH_2$ isotopologues with $n$ = 1–7. Although formally the frequency dependence has disappeared, and one could expect that the resulting $x^*$ values should equally well reflect the shape of $U(r)$ regardless of $n$, in reality the distance between the curves in Fig. 3c,d didn't change much: it decreased only slightly, compared to Fig. 3a,b. This means that the anharmonicity function $x^*(E)$ depends on the isotopologue mass in some indirect way.

It is noteworthy that due to the anharmonicity of the U($r$) potential, half of the fundamental transition energy does not coincide with the zero-point vibrational level energy, *i.e.* (G(1) – G(0))/2 – G(0) ≠ 0 (see Fig. S2). The decrease of this difference with increasing $n$ is caused by the fact that $\upsilon = 0$ и $\upsilon = 1$ levels approach the bottom of U($r$). In other words, the smaller region around the bottom of U($r$) is better described by a harmonic curve.

Finally, the obtained results allow us to address the question of the presence of a peculiar anomaly described in [1] – a sharp decrease of the absolute value of anharmonicity near the asymptote of the potential curve, based on theoretical and experimental data for the ground electronic states of $O_2$ and $Li_2$.[33] It was suggested that this is a manifestation of the change in the type of interatomic interaction at large internuclear distances from valence type to van der Waals type. However, Fig. 3 shows no abrupt changes of anharmonicity, which modulus continues to increase. It follows from Tables 2 and 3 that for some isotopologues the outer turning point of the last vibrational level reaches 6–7 Å, and the distance from this level to the asymptote is less than 1 $cm^{-1}$. In this region, the internuclear distance is an order of magnitude greater than $r_e$, and consequently the interaction between atoms is only a weak dispersive one. Thus, it can be assumed that the existence of the anharmonicity "spike" depends on the electronic state of the weakly interacting atoms. This question needs a special theoretical analysis.

## Conclusions

In this paper, on the basis of the ground electronic state term X $^1\Sigma_g^+$ of the hydrogen molecule for all seven known isotopologues $^nH_2$, $n$ = 1–7, we continue the development of the method of describing the potential curve of a diatomic molecule by the simplest Morse anharmonic model with a single anharmonicity coefficient $\omega_e x_e$. The necessity of introducing the second parameter, the bond energy $D_e$, leads to the ambiguity of the approximation, namely to the existence of two alternative solutions, M1($r$) and M2($r$), based on the choice of the initial parameters $\omega_e x_e$ (or $x_e$) or $D_e$, respectively. For simple terms, the potential function U($r$) lies between M1($r$) and M2($r$). For $^nH_2$, there are anomalies, when U($r$) intersects with M1($r$) or M2($r$).

To analyze the anomalies, the concept of systematic deviation $\delta(r)$ is introduced. At larger internuclear distances, for M1($r$) the deviation monotonously increases, reaching 1700–2770 $cm^{-1}$, so that the asymptote and several excess vibrational levels appear in the region of the continuous spectrum of U($r$). In the same region, systematic deviation of M2($r$) falls to zero, as M2($r$) and U($r$) share the same asymptote by definition.

The M2($r$) approximation of the $^1H_2$ potential function made in 1939 by Herzberg showed the presence of an intersection with the attraction branch around 1.5 Å, which was called the Herzberg anomaly.[1] Our data demonstrate its dependence on the isotopologue mass. For $^1H_2$, the intersection of M2($r$) with U($r$) occurs at ca. 1.252 Å, near the level $\upsilon$ = 3. Below this level, the M2($r$) curve is located anomalously with respect to U($r$), *i.e.* on the inner side of the potential well. In the region up to the second intersection point, the maximum U($r$) – M2($r$) difference is 125 $cm^{-1}$ at 1.081 Å and increases with increasing mass, reaching 142 $cm^{-1}$ for $^7H_2$ with the shift of intersection point to 1.269 Å.

For M1($r$), the Herzberg anomaly has a more complicated form. For $^1H_2$ in the region up to 1.9 Å, the systematic deviation has a wavy character: the maximum deviation is 447 $cm^{-1}$ at 1.298 Å, the minimum one is 186 $cm^{-1}$ at 1.822 Å and then it monotonously increases. With increasing mass, the deviation decreases, and for $^3H_2$ and subsequent isotopologues two intersections of the U($r$) curve with the M1($r$) pattern appear. Between these intersections, the M1($r$) curves evolve below U($r$). This region broadens with increasing isotope mass, and for $^7H_2$ it is in the interval 1.53–2.37 Å, roughly between the $\upsilon$ = 6 and $\upsilon$ = 12 levels, *i.e.* more than the upper half of the potential well.

Influence of the reduced mass on the vibrational structure of U($r$) is graphically illustrated in coordinates $\Delta_2G(\upsilon)$ *vs*. $\upsilon$, where $\upsilon$ is vibrational quantum number, and $\Delta_2G(\upsilon)$ is the second difference of functions $G(\upsilon)$, which also can be referred to as anharmonicity function. For $^nH_2$ isotopologues, there is an anomalous decrease in anharmonicity at the lower part of the potential with the formation of a characteristic convexity on the $\Delta_2G(\upsilon)$ *vs*. $\upsilon$ plot, which is presumably a manifestation of the Herzberg anomaly.[2,5] An increase in the reduced mass $\mu$ leads to a decrease in both anharmonicity and convexity due to a decrease in the region of the potential curve where $\Delta_2G(\upsilon)$ is measured.

The results obtained may be helpful for a wide range of specialists in various fields of chemical physics; a brief review of applications of the Morse function in recent years is given in paper [3]. The results are of particular value for teachers, since the description of the Morse function in textbooks and monographs is carried out according to the standard formed in the middle of the last century, and the question of the accuracy of approximation is scarcely addressed.

## Acknowledgments

The authors gratefully acknowledge the valuable discussions, criticism and advices from Professors I.G. Shenderovich, H.H. Limbach and I.G. Denisov. No financial support from funding organizations was used for this study.

*Supporting information for*

# Isotope effect in the Morse approximation of the ground state term of hydrogen molecule $^{n}H_2$, $n$ = 1÷7. Herzberg anomaly and anharmonicity

Gleb S. Denisov[a]; Edem R. Chakalov[b]; Peter M. Tolstoy[b*]

[a]Department of Physics, St. Petersburg State University, Russia

[b]Institute of Chemistry, St. Petersburg State University, Russia

[*]E-mail: peter.tolstoy@spbu.ru

Contents

Figure S1. Potential energy curve U(*r*) of the $^{1}H_2$ ground state X $^{1}\Sigma_g^+$ with vibrational levels and continuous term spectrum. Adapted from Fig. 48 in Herzberg, G. *Molecular Spectra and Molecular Structure I. Diatomic Molecules*; New York, 1939.

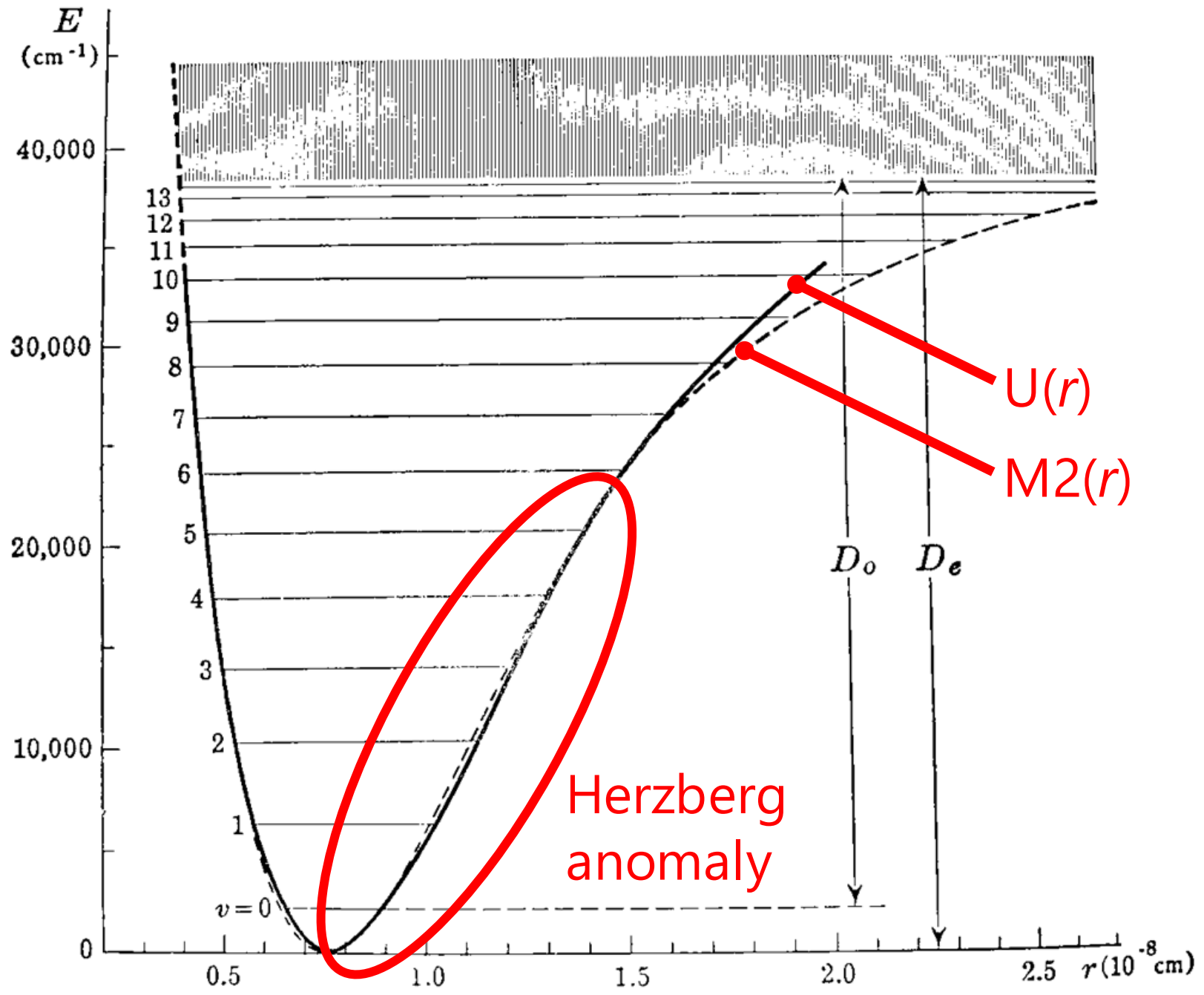

(the actual image above is adapted from the 1989's reprint of the 2nd edition: Herzberg, G. *Molecular Spectra and Molecular Structure I. Diatomic Molecules*; Van Nostrand, New York, 1950)

Table S1. Pachucki's potential energy curve U(*r*) of the hydrogen molecule in the ground state X $^1\Sigma_g^+$.

| *r*, **Å** | U(*r*), cm$^{-1}$ |
|---|---|
| 0.264589 | 142183.835817 |
| 0.269630 | 135908.984974 |
| 0.274671 | 129925.393278 |
| 0.279712 | 124217.390344 |
| 0.284753 | 118770.250429 |
| 0.289794 | 113570.136745 |
| 0.294835 | 108604.048494 |
| 0.299876 | 103859.770559 |
| 0.304917 | 99325.825730 |
| 0.309958 | 94991.429408 |
| 0.314999 | 90846.446687 |
| 0.320040 | 86881.351725 |
| 0.325081 | 83087.189329 |
| 0.330122 | 79455.538672 |
| 0.335163 | 75978.479043 |
| 0.340204 | 72648.557582 |
| 0.345245 | 69458.758878 |
| 0.350286 | 66402.476398 |
| 0.355327 | 63473.485627 |
| 0.360368 | 60665.918884 |
| 0.365409 | 57974.241703 |
| 0.370450 | 55393.230730 |
| 0.375491 | 52917.953065 |
| 0.380532 | 50543.746962 |
| 0.385573 | 48266.203832 |
| 0.390614 | 46081.151483 |
| 0.395655 | 43984.638524 |
| 0.400696 | 41972.919875 |
| 0.405737 | 40042.443316 |
| 0.410778 | 38189.837020 |
| 0.415819 | 36411.898007 |
| 0.420860 | 34705.581466 |
| 0.425901 | 33067.990876 |
| 0.430942 | 31496.368898 |
| 0.435983 | 29988.088955 |
| 0.441024 | 28540.647476 |
| 0.446065 | 27151.656737 |
| 0.451106 | 25818.838262 |
| 0.456147 | 24540.016740 |
| 0.461188 | 23313.114404 |
| 0.466229 | 22136.145849 |
| 0.471270 | 21007.213227 |
| 0.476311 | 19924.501808 |
| 0.481352 | 18886.275846 |
| 0.486393 | 17890.874739 |
| 0.491434 | 16936.709436 |
| 0.496475 | 16022.259076 |
| 0.501516 | 15146.067811 |
| 0.506557 | 14306.741821 |
| 0.511598 | 13502.946463 |
| 0.516639 | 12733.403561 |
| 0.521680 | 11996.888805 |
| 0.526721 | 11292.229247 |
| 0.531762 | 10618.300883 |
| 0.536803 | 9974.366296 |
| 0.541844 | 9358.983247 |
| 0.546885 | 8771.112102 |
| 0.551926 | 8209.837688 |
| 0.556967 | 7674.272960 |
| 0.562008 | 7163.558616 |
| 0.567049 | 6676.862686 |
| 0.572090 | 6213.380086 |
| 0.577131 | 5772.332144 |
| 0.582172 | 5352.966099 |
| 0.587213 | 4954.554570 |
| 0.592254 | 4576.395006 |
| 0.597295 | 4217.809106 |
| 0.602336 | 3878.142220 |
| 0.607377 | 3556.762732 |
| 0.612418 | 3253.061420 |
| 0.617459 | 2966.450804 |
| 0.622500 | 2696.364477 |
| 0.627541 | 2442.256424 |
| 0.632582 | 2203.600327 |
| 0.637623 | 1979.888870 |
| 0.642664 | 1770.633027 |
| 0.647705 | 1575.361353 |
| 0.652746 | 1393.619270 |
| 0.657787 | 1224.968352 |
| 0.662828 | 1068.985617 |
| 0.667869 | 925.262822 |
| 0.672910 | 793.405760 |
| 0.677951 | 673.033577 |
| 0.682992 | 563.778096 |
| 0.688034 | 465.283153 |
| 0.693075 | 377.231049 |
| 0.698116 | 299.231939 |
| 0.703157 | 230.980617 |
| 0.708198 | 172.168457 |
| 0.713239 | 122.495279 |
| 0.718280 | 81.669315 |
| 0.723321 | 49.407163 |
| 0.728362 | 25.433741 |
| 0.733403 | 9.482230 |
| 0.738444 | 1.294001 |
| 0.743485 | 0.618530 |
| 0.748526 | 7.213281 |
| 0.753567 | 20.843563 |
| 0.758608 | 41.282353 |
| 0.763649 | 68.310078 |
| 0.768690 | 101.723299 |
| 0.773731 | 141.303153 |
| 0.778772 | 186.850460 |
| 0.783813 | 238.172219 |
| 0.788854 | 295.080937 |
| 0.793895 | 357.394488 |
| 0.798936 | 424.935973 |
| 0.803977 | 497.533586 |
| 0.809018 | 575.020477 |
| 0.814059 | 657.234616 |
| 0.819100 | 744.018662 |
| 0.824141 | 835.219828 |
| 0.829182 | 930.689754 |
| 0.834223 | 1030.284378 |
| 0.839264 | 1133.863807 |
| 0.844305 | 1241.292198 |
| 0.849346 | 1352.437630 |
| 0.854387 | 1467.171988 |
| 0.859428 | 1585.370846 |
| 0.864469 | 1706.913351 |
| 0.869510 | 1831.682110 |
| 0.874551 | 1959.563085 |
| 0.879592 | 2090.445481 |
| 0.884633 | 2224.221646 |
| 0.889674 | 2360.786969 |
| 0.894715 | 2500.039784 |
| 0.899756 | 2641.881270 |
| 0.904797 | 2786.215366 |
| 0.909838 | 2932.948676 |
| 0.914879 | 3081.990389 |
| 0.919920 | 3233.252189 |
| 0.924961 | 3386.648183 |
| 0.930002 | 3542.094817 |
| 0.935043 | 3699.510807 |
| 0.940084 | 3858.817064 |
| 0.945125 | 4019.936630 |
| 0.950166 | 4182.794608 |
| 0.955207 | 4347.318103 |
| 0.960248 | 4513.436159 |
| 0.965289 | 4681.079702 |
| 0.970330 | 4850.181486 |
| 0.975371 | 5020.676038 |
| 0.980412 | 5192.499612 |
| 0.985453 | 5365.590134 |
| 0.990494 | 5539.887160 |
| 0.995535 | 5715.331831 |
| 1.000576 | 5891.866831 |
| 1.005617 | 6069.436344 |
| 1.010658 | 6248.015579 |
| 1.015699 | 6427.519435 |
| 1.020740 | 6607.887322 |
| 1.025781 | 6789.071536 |
| 1.030822 | 6971.025373 |
| 1.035863 | 7153.703118 |
| 1.040904 | 7337.060033 |
| 1.045945 | 7521.052346 |
| 1.050986 | 7705.637244 |
| 1.056027 | 7890.772859 |
| 1.061068 | 8076.418259 |
| 1.066109 | 8262.533431 |
| 1.071150 | 8449.079276 |
| 1.076191 | 8636.017590 |
| 1.081232 | 8823.311058 |
| 1.086273 | 9010.923235 |
| 1.091314 | 9198.818534 |
| 1.096355 | 9386.962217 |
| 1.101396 | 9575.320376 |
| 1.106437 | 9763.859922 |
| 1.111479 | 9952.548570 |
| 1.116520 | 10141.354827 |
| 1.121561 | 10330.247975 |
| 1.126602 | 10519.198056 |
| 1.131643 | 10708.175863 |
| 1.136684 | 10897.152920 |
| 1.141725 | 11086.101469 |
| 1.146766 | 11274.994456 |
| 1.151807 | 11463.805517 |
| 1.156848 | 11652.508963 |
| 1.161889 | 11841.079764 |
| 1.166930 | 12029.493535 |
| 1.171971 | 12217.726526 |
| 1.177012 | 12405.755599 |
| 1.182053 | 12593.558223 |
| 1.187094 | 12781.112453 |
| 1.192135 | 12968.396919 |
| 1.197176 | 13155.390812 |
| 1.202217 | 13342.073870 |
| 1.207258 | 13528.426365 |
| 1.212299 | 13714.429088 |
| 1.217340 | 13900.063337 |
| 1.222381 | 14085.310906 |
| 1.227422 | 14270.154069 |
| 1.232463 | 14454.575569 |
| 1.237504 | 14638.558605 |
| 1.242545 | 14822.086824 |
| 1.247586 | 15005.144303 |
| 1.252627 | 15187.715542 |
| 1.257668 | 15369.785452 |
| 1.262709 | 15551.339343 |
| 1.267750 | 15732.362917 |
| 1.272791 | 15912.842252 |
| 1.277832 | 16092.763798 |
| 1.282873 | 16272.114363 |
| 1.287914 | 16450.881108 |
| 1.292955 | 16629.051533 |
| 1.297996 | 16806.613474 |
| 1.303037 | 16983.555090 |
| 1.308078 | 17159.864857 |
| 1.313119 | 17335.531561 |
| 1.318160 | 17510.544290 |
| 1.323201 | 17684.892427 |
| 1.328242 | 17858.565643 |
| 1.333283 | 18031.553893 |
| 1.338324 | 18203.847405 |
| 1.343365 | 18375.436680 |
| 1.348406 | 18546.312481 |
| 1.353447 | 18716.465832 |
| 1.358488 | 18885.888013 |
| 1.363529 | 19054.570551 |
| 1.368570 | 19222.505222 |
| 1.373611 | 19389.684039 |
| 1.378652 | 19556.099257 |
| 1.383693 | 19721.743360 |
| 1.388734 | 19886.609065 |
| 1.393775 | 20050.689316 |
| 1.398816 | 20213.977278 |
| 1.403857 | 20376.466338 |
| 1.408898 | 20538.150099 |
| 1.413939 | 20699.022381 |
| 1.418980 | 20859.077213 |
| 1.424021 | 21018.308835 |
| 1.429062 | 21176.711693 |
| 1.434103 | 21334.280438 |
| 1.439144 | 21491.009921 |
| 1.444185 | 21646.895193 |
| 1.449226 | 21801.931503 |
| 1.454267 | 21956.114292 |
| 1.459308 | 22109.439193 |
| 1.464349 | 22261.902031 |
| 1.469390 | 22413.498814 |
| 1.474431 | 22564.225734 |
| 1.479472 | 22714.079168 |
| 1.484513 | 22863.055666 |
| 1.489554 | 23011.135713 |
| 1.494595 | 23158.355065 |
| 1.499636 | 23304.687899 |
| 1.504677 | 23450.131470 |
| 1.509718 | 23594.683179 |
| 1.514759 | 23738.340561 |
| 1.519800 | 23881.101291 |
| 1.524841 | 24022.963185 |
| 1.529883 | 24163.924185 |
| 1.534924 | 24303.982372 |
| 1.539965 | 24443.135960 |
| 1.545006 | 24581.383293 |
| 1.550047 | 24718.722846 |
| 1.555088 | 24855.153221 |
| 1.560129 | 24990.673150 |
| 1.565170 | 25125.281489 |
| 1.570211 | 25258.977218 |
| 1.575252 | 25391.759444 |
| 1.580293 | 25523.627392 |
| 1.585334 | 25654.580411 |
| 1.590375 | 25784.617966 |
| 1.595416 | 25913.739644 |
| 1.600457 | 26041.945147 |
| 1.605498 | 26169.234290 |
| 1.610539 | 26295.607006 |
| 1.615580 | 26421.063339 |
| 1.620621 | 26545.603443 |
| 1.625662 | 26669.227585 |
| 1.630703 | 26791.936138 |
| 1.635744 | 26913.729583 |
| 1.640785 | 27034.608507 |
| 1.645826 | 27154.573604 |
| 1.650867 | 27273.625666 |
| 1.655908 | 27391.765593 |
| 1.660949 | 27508.994380 |
| 1.665990 | 27625.313126 |
| 1.671031 | 27740.723025 |
| 1.676072 | 27855.225368 |
| 1.681113 | 27968.821541 |
| 1.686154 | 28081.513026 |
| 1.691195 | 28193.301394 |
| 1.696236 | 28304.188311 |
| 1.701277 | 28414.175530 |
| 1.706318 | 28523.264894 |
| 1.711359 | 28631.458332 |
| 1.716400 | 28738.757861 |
| 1.721441 | 28845.165580 |
| 1.726482 | 28950.683674 |
| 1.731523 | 29055.314407 |
| 1.736564 | 29159.060127 |
| 1.741605 | 29261.923259 |
| 1.746646 | 29363.906306 |
| 1.751687 | 29465.011850 |
| 1.756728 | 29565.242546 |
| 1.761769 | 29664.601124 |
| 1.766810 | 29763.090389 |
| 1.771851 | 29860.713213 |
| 1.776892 | 29957.472544 |
| 1.781933 | 30053.371394 |
| 1.786974 | 30148.412847 |
| 1.792015 | 30242.600051 |
| 1.797056 | 30335.936220 |
| 1.802097 | 30428.424632 |
| 1.807138 | 30520.068628 |
| 1.812179 | 30610.871611 |
| 1.817220 | 30700.837043 |
| 1.822261 | 30789.968448 |
| 1.827302 | 30878.269403 |
| 1.832343 | 30965.743547 |
| 1.837384 | 31052.394570 |
| 1.842425 | 31138.226220 |
| 1.847466 | 31223.242295 |
| 1.852507 | 31307.446646 |
| 1.857548 | 31390.843175 |
| 1.862589 | 31473.435833 |
| 1.867630 | 31555.228618 |
| 1.872671 | 31636.225576 |
| 1.877712 | 31716.430800 |
| 1.882753 | 31795.848425 |
| 1.887794 | 31874.482632 |
| 1.892835 | 31952.337642 |
| 1.897876 | 32029.417718 |
| 1.902917 | 32105.727163 |
| 1.907958 | 32181.270318 |
| 1.912999 | 32256.051564 |
| 1.918040 | 32330.075314 |
| 1.923081 | 32403.346020 |
| 1.928122 | 32475.868165 |
| 1.933163 | 32547.646269 |
| 1.938204 | 32618.684879 |
| 1.943245 | 32688.988575 |
| 1.948286 | 32758.561967 |
| 1.953328 | 32827.409691 |
| 1.958369 | 32895.536413 |
| 1.963410 | 32962.946823 |
| 1.968451 | 33029.645636 |
| 1.973492 | 33095.637591 |
| 1.978533 | 33160.927450 |
| 1.983574 | 33225.519996 |
| 1.988615 | 33289.420032 |
| 1.993656 | 33352.632381 |
| 1.998697 | 33415.161884 |
| 2.003738 | 33477.013398 |
| 2.008779 | 33538.191798 |
| 2.013820 | 33598.701971 |
| 2.018861 | 33658.548820 |
| 2.023902 | 33717.737259 |
| 2.028943 | 33776.272215 |
| 2.033984 | 33834.158624 |
| 2.039025 | 33891.401433 |
| 2.044066 | 33948.005596 |
| 2.049107 | 34003.976074 |
| 2.054148 | 34059.317835 |
| 2.059189 | 34114.035852 |
| 2.064230 | 34168.135103 |
| 2.069271 | 34221.622527 |
| 2.074312 | 34274.500186 |
| 2.079353 | 34326.773660 |
| 2.084394 | 34378.447977 |
| 2.089435 | 34429.528165 |
| 2.094476 | 34480.019243 |
| 2.099517 | 34529.926225 |
| 2.104558 | 34579.254119 |
| 2.109599 | 34628.007927 |
| 2.114640 | 34676.192639 |
| 2.119681 | 34723.813240 |
| 2.124722 | 34770.874704 |
| 2.129763 | 34817.381996 |
| 2.134804 | 34863.340069 |
| 2.139845 | 34908.753865 |
| 2.144886 | 34953.628314 |
| 2.149927 | 34997.968335 |
| 2.154968 | 35041.778832 |
| 2.160009 | 35085.064696 |
| 2.165050 | 35127.830802 |
| 2.170091 | 35170.082013 |
| 2.175132 | 35211.823173 |
| 2.180173 | 35253.059114 |
| 2.185214 | 35293.794647 |
| 2.190255 | 35334.034568 |
| 2.195296 | 35373.783657 |
| 2.200337 | 35413.046672 |
| 2.205378 | 35451.828354 |
| 2.210419 | 35490.133426 |
| 2.215460 | 35527.966590 |
| 2.220501 | 35565.332526 |
| 2.225542 | 35602.235898 |
| 2.230583 | 35638.681343 |
| 2.235624 | 35674.673482 |
| 2.240665 | 35710.216910 |
| 2.245706 | 35745.316200 |
| 2.250747 | 35779.975906 |
| 2.255788 | 35814.200553 |
| 2.260829 | 35847.994646 |
| 2.265870 | 35881.362665 |
| 2.270911 | 35914.309066 |
| 2.275952 | 35946.838280 |
| 2.280993 | 35978.954712 |
| 2.286034 | 36010.662742 |
| 2.291075 | 36041.966726 |
| 2.296116 | 36072.870992 |
| 2.301157 | 36103.379841 |
| 2.306198 | 36133.497548 |
| 2.311239 | 36163.228363 |
| 2.316280 | 36192.576506 |
| 2.321321 | 36221.546170 |
| 2.326362 | 36250.141522 |
| 2.331403 | 36278.366698 |
| 2.336444 | 36306.225808 |
| 2.341485 | 36333.722933 |
| 2.346526 | 36360.862126 |
| 2.351567 | 36387.647408 |
| 2.356608 | 36414.082775 |
| 2.361649 | 36440.172192 |
| 2.366690 | 36465.919593 |
| 2.371732 | 36491.328885 |
| 2.376773 | 36516.403943 |
| 2.381814 | 36541.148614 |
| 2.386855 | 36565.566713 |
| 2.391896 | 36589.662028 |
| 2.396937 | 36613.438312 |
| 2.401978 | 36636.899291 |
| 2.407019 | 36660.048661 |
| 2.412060 | 36682.890084 |
| 2.417101 | 36705.427195 |
| 2.422142 | 36727.663597 |
| 2.427183 | 36749.602861 |
| 2.432224 | 36771.248528 |
| 2.437265 | 36792.604109 |
| 2.442306 | 36813.673083 |
| 2.447347 | 36834.458899 |
| 2.452388 | 36854.964974 |
| 2.457429 | 36875.194694 |
| 2.462470 | 36895.151416 |
| 2.467511 | 36914.838464 |
| 2.472552 | 36934.259131 |
| 2.477593 | 36953.416680 |
| 2.482634 | 36972.314344 |
| 2.487675 | 36990.955324 |
| 2.492716 | 37009.342789 |
| 2.497757 | 37027.479880 |
| 2.502798 | 37045.369706 |
| 2.507839 | 37063.015344 |
| 2.512880 | 37080.419843 |
| 2.517921 | 37097.586220 |
| 2.522962 | 37114.517462 |
| 2.528003 | 37131.216527 |
| 2.533044 | 37147.686340 |
| 2.538085 | 37163.929798 |
| 2.543126 | 37179.949768 |
| 2.548167 | 37195.749088 |
| 2.553208 | 37211.330563 |
| 2.558249 | 37226.696972 |
| 2.563290 | 37241.851063 |
| 2.568331 | 37256.795555 |
| 2.573372 | 37271.533137 |
| 2.578413 | 37286.066471 |
| 2.583454 | 37300.398189 |
| 2.588495 | 37314.530894 |
| 2.593536 | 37328.467161 |
| 2.598577 | 37342.209536 |
| 2.603618 | 37355.760538 |
| 2.608659 | 37369.122658 |
| 2.613700 | 37382.298357 |
| 2.618741 | 37395.290072 |
| 2.623782 | 37408.100209 |
| 2.628823 | 37420.731150 |
| 2.633864 | 37433.185247 |
| 2.638905 | 37445.464827 |
| 2.643946 | 37457.572190 |
| 2.648987 | 37469.509610 |
| 2.654028 | 37481.279334 |
| 2.659069 | 37492.883583 |
| 2.664110 | 37504.324553 |
| 2.669151 | 37515.604413 |
| 2.674192 | 37526.725308 |
| 2.679233 | 37537.689358 |
| 2.684274 | 37548.498657 |
| 2.689315 | 37559.155275 |
| 2.694356 | 37569.661257 |
| 2.699397 | 37580.018623 |
| 2.704438 | 37590.229372 |
| 2.709479 | 37600.295475 |
| 2.714520 | 37610.218882 |
| 2.719561 | 37620.001520 |
| 2.724602 | 37629.645290 |
| 2.729643 | 37639.152074 |
| 2.734684 | 37648.523728 |
| 2.739725 | 37657.762086 |
| 2.744766 | 37666.868963 |
| 2.749807 | 37675.846146 |
| 2.754848 | 37684.695406 |
| 2.759889 | 37693.418489 |
| 2.764930 | 37702.017121 |
| 2.769971 | 37710.493005 |
| 2.775012 | 37718.847826 |
| 2.780053 | 37727.083245 |
| 2.785094 | 37735.200906 |
| 2.790135 | 37743.202429 |
| 2.795177 | 37751.089416 |
| 2.800218 | 37758.863450 |
| 2.805259 | 37766.526091 |
| 2.810300 | 37774.078884 |
| 2.815341 | 37781.523350 |
| 2.820382 | 37788.860995 |
| 2.825423 | 37796.093302 |
| 2.830464 | 37803.221740 |
| 2.835505 | 37810.247756 |
| 2.840546 | 37817.172780 |
| 2.845587 | 37823.998224 |
| 2.850628 | 37830.725480 |
| 2.855669 | 37837.355926 |
| 2.860710 | 37843.890920 |
| 2.865751 | 37850.329571 |
| 2.870792 | 37856.679519 |
| 2.875833 | 37862.937965 |
| 2.880874 | 37869.106191 |
| 2.885915 | 37875.185460 |
| 2.890956 | 37881.177019 |
| 2.895997 | 37887.082099 |
| 2.901038 | 37892.901915 |
| 2.906079 | 37898.637667 |
| 2.911120 | 37904.290537 |
| 2.916161 | 37909.861695 |
| 2.921202 | 37915.352293 |
| 2.926243 | 37920.763469 |
| 2.931284 | 37926.096346 |
| 2.936325 | 37931.352032 |
| 2.941366 | 37936.531620 |
| 2.946407 | 37941.636190 |
| 2.951448 | 37946.666805 |
| 2.956489 | 37951.624518 |
| 2.961530 | 37956.510363 |
| 2.966571 | 37961.325365 |
| 2.971612 | 37966.070532 |
| 2.976653 | 37970.746859 |
| 2.981694 | 37975.355330 |
| 2.986735 | 37979.896914 |
| 2.991776 | 37984.372565 |
| 2.996817 | 37988.783229 |
| 3.001858 | 37993.129834 |
| 3.006899 | 37997.413299 |
| 3.011940 | 38001.634530 |
| 3.016981 | 38005.794420 |
| 3.022022 | 38009.893850 |
| 3.027063 | 38013.933688 |
| 3.032104 | 38017.914792 |
| 3.037145 | 38021.838007 |
| 3.042186 | 38025.704167 |
| 3.047227 | 38029.514095 |
| 3.052268 | 38033.268602 |
| 3.057309 | 38036.968488 |
| 3.062350 | 38040.614541 |
| 3.067391 | 38044.207539 |
| 3.072432 | 38047.748251 |
| 3.077473 | 38051.237432 |
| 3.082514 | 38054.675829 |
| 3.087555 | 38058.064177 |
| 3.092596 | 38061.403202 |
| 3.097637 | 38064.693619 |
| 3.102678 | 38067.936133 |
| 3.107719 | 38071.131441 |
| 3.112760 | 38074.280228 |
| 3.117801 | 38077.383169 |
| 3.122842 | 38080.440932 |
| 3.127883 | 38083.454175 |
| 3.132924 | 38086.423543 |
| 3.137965 | 38089.349678 |
| 3.143006 | 38092.233208 |
| 3.148047 | 38095.074754 |
| 3.153088 | 38097.874929 |
| 3.158129 | 38100.634335 |
| 3.163170 | 38103.353568 |
| 3.168211 | 38106.033214 |
| 3.173252 | 38108.673850 |
| 3.178293 | 38111.276047 |
| 3.183334 | 38113.840367 |
| 3.188375 | 38116.367362 |
| 3.193416 | 38118.857578 |
| 3.198457 | 38121.311553 |
| 3.203498 | 38123.729818 |
| 3.208539 | 38126.112894 |
| 3.213581 | 38128.461296 |
| 3.218622 | 38130.775533 |
| 3.223663 | 38133.056103 |
| 3.228704 | 38135.303501 |
| 3.233745 | 38137.518212 |
| 3.238786 | 38139.700714 |
| 3.243827 | 38141.851480 |
| 3.248868 | 38143.970974 |
| 3.253909 | 38146.059656 |
| 3.258950 | 38148.117976 |
| 3.263991 | 38150.146379 |
| 3.269032 | 38152.145305 |
| 3.274073 | 38154.115184 |
| 3.279114 | 38156.056444 |
| 3.284155 | 38157.969504 |
| 3.289196 | 38159.854777 |
| 3.294237 | 38161.712670 |
| 3.299278 | 38163.543586 |
| 3.304319 | 38165.347921 |
| 3.309360 | 38167.126063 |
| 3.314401 | 38168.878397 |
| 3.319442 | 38170.605302 |
| 3.324483 | 38172.307150 |
| 3.329524 | 38173.984309 |
| 3.334565 | 38175.637142 |
| 3.339606 | 38177.266004 |
| 3.344647 | 38178.871248 |
| 3.349688 | 38180.453220 |
| 3.354729 | 38182.012261 |
| 3.359770 | 38183.548708 |
| 3.364811 | 38185.062891 |
| 3.369852 | 38186.555138 |
| 3.374893 | 38188.025769 |
| 3.379934 | 38189.475102 |
| 3.384975 | 38190.903449 |
| 3.390016 | 38192.311118 |
| 3.395057 | 38193.698411 |
| 3.400098 | 38195.065628 |
| 3.405139 | 38196.413062 |
| 3.410180 | 38197.741003 |
| 3.415221 | 38199.049737 |
| 3.420262 | 38200.339546 |
| 3.425303 | 38201.610706 |
| 3.430344 | 38202.863491 |
| 3.435385 | 38204.098170 |
| 3.440426 | 38205.315008 |
| 3.445467 | 38206.514266 |
| 3.450508 | 38207.696202 |
| 3.455549 | 38208.861069 |
| 3.460590 | 38210.009118 |
| 3.465631 | 38211.140594 |
| 3.470672 | 38212.255739 |
| 3.475713 | 38213.354794 |
| 3.480754 | 38214.437993 |
| 3.485795 | 38215.505568 |
| 3.490836 | 38216.557748 |
| 3.495877 | 38217.594759 |
| 3.500918 | 38218.616822 |
| 3.505959 | 38219.624155 |
| 3.511000 | 38220.616975 |
| 3.516041 | 38221.595493 |
| 3.521082 | 38222.559919 |
| 3.526123 | 38223.510459 |
| 3.531164 | 38224.447316 |
| 3.536205 | 38225.370690 |
| 3.541246 | 38226.280778 |
| 3.546287 | 38227.177775 |
| 3.551328 | 38228.061871 |
| 3.556369 | 38228.933257 |
| 3.561410 | 38229.792116 |
| 3.566451 | 38230.638634 |
| 3.571492 | 38231.472990 |
| 3.576533 | 38232.295362 |
| 3.581574 | 38233.105925 |
| 3.586615 | 38233.904853 |
| 3.591656 | 38234.692315 |
| 3.596697 | 38235.468479 |
| 3.601738 | 38236.233511 |
| 3.606779 | 38236.987573 |
| 3.611820 | 38237.730826 |
| 3.616861 | 38238.463428 |
| 3.621902 | 38239.185535 |
| 3.626943 | 38239.897301 |
| 3.631984 | 38240.598878 |
| 3.637026 | 38241.290414 |
| 3.642067 | 38241.972056 |
| 3.647108 | 38242.643951 |
| 3.652149 | 38243.306239 |
| 3.657190 | 38243.959063 |
| 3.662231 | 38244.602562 |
| 3.667272 | 38245.236871 |
| 3.672313 | 38245.862126 |
| 3.677354 | 38246.478460 |
| 3.682395 | 38247.086003 |
| 3.687436 | 38247.684886 |
| 3.692477 | 38248.275234 |
| 3.697518 | 38248.857174 |
| 3.702559 | 38249.430829 |
| 3.707600 | 38249.996321 |
| 3.712641 | 38250.553770 |
| 3.717682 | 38251.103295 |
| 3.722723 | 38251.645012 |
| 3.727764 | 38252.179036 |
| 3.732805 | 38252.705481 |
| 3.737846 | 38253.224458 |
| 3.742887 | 38253.736078 |
| 3.747928 | 38254.240449 |
| 3.752969 | 38254.737678 |
| 3.758010 | 38255.227870 |
| 3.763051 | 38255.711131 |
| 3.768092 | 38256.187561 |
| 3.773133 | 38256.657263 |
| 3.778174 | 38257.120335 |
| 3.783215 | 38257.576877 |
| 3.788256 | 38258.026984 |
| 3.793297 | 38258.470752 |
| 3.798338 | 38258.908276 |
| 3.803379 | 38259.339647 |
| 3.808420 | 38259.764958 |
| 3.813461 | 38260.184297 |
| 3.818502 | 38260.597755 |
| 3.823543 | 38261.005418 |
| 3.828584 | 38261.407372 |
| 3.833625 | 38261.803703 |
| 3.838666 | 38262.194494 |
| 3.843707 | 38262.579828 |
| 3.848748 | 38262.959787 |
| 3.853789 | 38263.334449 |
| 3.858830 | 38263.703895 |
| 3.863871 | 38264.068202 |
| 3.868912 | 38264.427446 |
| 3.873953 | 38264.781704 |
| 3.878994 | 38265.131051 |
| 3.884035 | 38265.475558 |
| 3.889076 | 38265.815299 |
| 3.894117 | 38266.150345 |
| 3.899158 | 38266.480766 |
| 3.904199 | 38266.806632 |
| 3.909240 | 38267.128010 |
| 3.914281 | 38267.444968 |
| 3.919322 | 38267.757571 |
| 3.924363 | 38268.065886 |
| 3.929404 | 38268.369975 |
| 3.934445 | 38268.669904 |
| 3.939486 | 38268.965732 |
| 3.944527 | 38269.257523 |
| 3.949568 | 38269.545337 |
| 3.954609 | 38269.829232 |
| 3.959650 | 38270.109268 |
| 3.964691 | 38270.385502 |
| 3.969732 | 38270.657991 |
| 3.974773 | 38270.946228 |
| 3.979814 | 38271.209324 |
| 3.984855 | 38271.469006 |
| 3.989896 | 38271.725319 |
| 3.994937 | 38271.978304 |
| 3.999978 | 38272.228002 |
| 4.005019 | 38272.474454 |
| 4.010060 | 38272.717702 |
| 4.015101 | 38272.957785 |
| 4.020142 | 38273.194743 |
| 4.025183 | 38273.428615 |
| 4.030224 | 38273.659441 |
| 4.035265 | 38273.887259 |
| 4.040306 | 38274.112107 |
| 4.045347 | 38274.334022 |
| 4.050388 | 38274.553043 |
| 4.055430 | 38274.769205 |
| 4.060471 | 38274.982546 |
| 4.065512 | 38275.193102 |
| 4.070553 | 38275.400907 |
| 4.075594 | 38275.605999 |
| 4.080635 | 38275.808410 |
| 4.085676 | 38276.008177 |
| 4.090717 | 38276.205332 |
| 4.095758 | 38276.399911 |
| 4.100799 | 38276.591945 |
| 4.105840 | 38276.781469 |
| 4.110881 | 38276.968515 |
| 4.115922 | 38277.153115 |
| 4.120963 | 38277.335301 |
| 4.126004 | 38277.515105 |
| 4.131045 | 38277.692558 |
| 4.136086 | 38277.867690 |
| 4.141127 | 38278.040533 |
| 4.146168 | 38278.211116 |
| 4.151209 | 38278.379469 |
| 4.156250 | 38278.545621 |
| 4.161291 | 38278.709603 |
| 4.166332 | 38278.871441 |
| 4.171373 | 38279.031166 |
| 4.176414 | 38279.188805 |
| 4.181455 | 38279.344385 |
| 4.186496 | 38279.497935 |
| 4.191537 | 38279.649481 |
| 4.196578 | 38279.799050 |
| 4.201619 | 38279.946669 |
| 4.206660 | 38280.092364 |
| 4.211701 | 38280.236161 |
| 4.216742 | 38280.378084 |
| 4.221783 | 38280.518161 |
| 4.226824 | 38280.656414 |
| 4.231865 | 38280.792870 |
| 4.236906 | 38280.927553 |
| 4.241947 | 38281.060486 |
| 4.246988 | 38281.191693 |
| 4.252029 | 38281.321198 |
| 4.257070 | 38281.449024 |
| 4.262111 | 38281.575193 |
| 4.267152 | 38281.699729 |
| 4.272193 | 38281.822654 |
| 4.277234 | 38281.943990 |
| 4.282275 | 38282.063758 |
| 4.287316 | 38282.181981 |
| 4.292357 | 38282.298678 |
| 4.297398 | 38282.413872 |
| 4.302439 | 38282.527584 |
| 4.307480 | 38282.639832 |
| 4.312521 | 38282.750639 |
| 4.317562 | 38282.860022 |
| 4.322603 | 38282.968004 |
| 4.327644 | 38283.074602 |
| 4.332685 | 38283.179835 |
| 4.337726 | 38283.283724 |
| 4.342767 | 38283.386286 |
| 4.347808 | 38283.487540 |
| 4.352849 | 38283.587504 |
| 4.357890 | 38283.686197 |
| 4.362931 | 38283.783635 |
| 4.367972 | 38283.879836 |
| 4.373013 | 38283.974818 |
| 4.378054 | 38284.068598 |
| 4.383095 | 38284.161192 |
| 4.388136 | 38284.252618 |
| 4.393177 | 38284.342890 |
| 4.398218 | 38284.432027 |
| 4.403259 | 38284.520042 |
| 4.408300 | 38284.606954 |
| 4.413341 | 38284.692775 |
| 4.418382 | 38284.777523 |
| 4.423423 | 38284.861212 |
| 4.428464 | 38284.943857 |
| 4.433505 | 38285.025473 |
| 4.438546 | 38285.106074 |
| 4.443587 | 38285.185674 |
| 4.448628 | 38285.264288 |
| 4.453669 | 38285.341929 |
| 4.458710 | 38285.418612 |
| 4.463751 | 38285.494349 |
| 4.468792 | 38285.569154 |
| 4.473833 | 38285.643040 |
| 4.478875 | 38285.716021 |
| 4.483916 | 38285.788108 |
| 4.488957 | 38285.859315 |
| 4.493998 | 38285.929653 |
| 4.499039 | 38285.999135 |
| 4.504080 | 38286.067774 |
| 4.509121 | 38286.135581 |
| 4.514162 | 38286.202567 |
| 4.519203 | 38286.268745 |
| 4.524244 | 38286.334125 |
| 4.529285 | 38286.398719 |
| 4.534326 | 38286.462538 |
| 4.539367 | 38286.525593 |
| 4.544408 | 38286.587895 |
| 4.549449 | 38286.649454 |
| 4.554490 | 38286.710280 |
| 4.559531 | 38286.770384 |
| 4.564572 | 38286.829777 |
| 4.569613 | 38286.888467 |
| 4.574654 | 38286.946465 |
| 4.579695 | 38287.003780 |
| 4.584736 | 38287.060423 |
| 4.589777 | 38287.116402 |
| 4.594818 | 38287.171726 |
| 4.599859 | 38287.226406 |
| 4.604900 | 38287.280449 |
| 4.609941 | 38287.333865 |
| 4.614982 | 38287.386662 |
| 4.620023 | 38287.438849 |
| 4.625064 | 38287.490434 |
| 4.630105 | 38287.541426 |
| 4.635146 | 38287.591833 |
| 4.640187 | 38287.641663 |
| 4.645228 | 38287.690924 |
| 4.650269 | 38287.739623 |
| 4.655310 | 38287.787769 |
| 4.660351 | 38287.835369 |
| 4.665392 | 38287.882431 |
| 4.670433 | 38287.928961 |
| 4.675474 | 38287.974968 |
| 4.680515 | 38288.020457 |
| 4.685556 | 38288.065437 |
| 4.690597 | 38288.109914 |
| 4.695638 | 38288.153895 |
| 4.700679 | 38288.197387 |
| 4.705720 | 38288.240396 |
| 4.710761 | 38288.282928 |
| 4.715802 | 38288.324991 |
| 4.720843 | 38288.366590 |
| 4.725884 | 38288.407732 |
| 4.730925 | 38288.448423 |
| 4.735966 | 38288.488668 |
| 4.741007 | 38288.528474 |
| 4.746048 | 38288.567846 |
| 4.751089 | 38288.606790 |
| 4.756130 | 38288.645312 |
| 4.761171 | 38288.683417 |
| 4.766212 | 38288.721111 |
| 4.771253 | 38288.758400 |
| 4.776294 | 38288.795287 |
| 4.781335 | 38288.831779 |
| 4.786376 | 38288.867881 |
| 4.791417 | 38288.903597 |
| 4.796458 | 38288.938933 |
| 4.801499 | 38288.973894 |
| 4.806540 | 38289.008483 |
| 4.811581 | 38289.042707 |
| 4.816622 | 38289.076570 |
| 4.821663 | 38289.110076 |
| 4.826704 | 38289.143229 |
| 4.831745 | 38289.176035 |
| 4.836786 | 38289.208497 |
| 4.841827 | 38289.240620 |
| 4.846868 | 38289.272408 |
| 4.851909 | 38289.303865 |
| 4.856950 | 38289.334996 |
| 4.861991 | 38289.365803 |
| 4.867032 | 38289.396292 |
| 4.872073 | 38289.426466 |
| 4.877114 | 38289.456329 |
| 4.882155 | 38289.485885 |
| 4.887196 | 38289.515137 |
| 4.892237 | 38289.544090 |
| 4.897278 | 38289.572746 |
| 4.902320 | 38289.601109 |
| 4.907361 | 38289.629183 |
| 4.912402 | 38289.656971 |
| 4.917443 | 38289.684477 |
| 4.922484 | 38289.711703 |
| 4.927525 | 38289.738654 |
| 4.932566 | 38289.765332 |
| 4.937607 | 38289.791741 |
| 4.942648 | 38289.817883 |
| 4.947689 | 38289.843762 |
| 4.952730 | 38289.869381 |
| 4.957771 | 38289.894743 |
| 4.962812 | 38289.919850 |
| 4.967853 | 38289.944706 |
| 4.972894 | 38289.969313 |
| 4.977935 | 38289.993675 |
| 4.982976 | 38290.017794 |
| 4.988017 | 38290.041673 |
| 4.993058 | 38290.065314 |
| 4.998099 | 38290.088721 |
| 5.003140 | 38290.111895 |
| 5.008181 | 38290.134839 |
| 5.013222 | 38290.157557 |
| 5.018263 | 38290.180049 |
| 5.023304 | 38290.202320 |
| 5.028345 | 38290.224371 |
| 5.033386 | 38290.246205 |
| 5.038427 | 38290.267824 |
| 5.043468 | 38290.289230 |
| 5.048509 | 38290.310426 |
| 5.053550 | 38290.331414 |
| 5.058591 | 38290.352196 |
| 5.063632 | 38290.372775 |
| 5.068673 | 38290.393152 |
| 5.073714 | 38290.413331 |
| 5.078755 | 38290.433312 |
| 5.083796 | 38290.453098 |
| 5.088837 | 38290.472691 |
| 5.093878 | 38290.492094 |
| 5.098919 | 38290.511308 |
| 5.103960 | 38290.530335 |
| 5.109001 | 38290.549177 |
| 5.114042 | 38290.567836 |
| 5.119083 | 38290.586315 |
| 5.124124 | 38290.604614 |
| 5.129165 | 38290.622736 |
| 5.134206 | 38290.640683 |
| 5.139247 | 38290.658457 |
| 5.144288 | 38290.676059 |
| 5.149329 | 38290.693492 |
| 5.154370 | 38290.710756 |
| 5.159411 | 38290.727854 |
| 5.164452 | 38290.744788 |
| 5.169493 | 38290.761559 |
| 5.174534 | 38290.778168 |
| 5.179575 | 38290.794619 |
| 5.184616 | 38290.810912 |
| 5.189657 | 38290.827049 |
| 5.194698 | 38290.843032 |
| 5.199739 | 38290.858861 |
| 5.204780 | 38290.874540 |
| 5.209821 | 38290.890070 |
| 5.214862 | 38290.905451 |
| 5.219903 | 38290.920686 |
| 5.224944 | 38290.935777 |
| 5.229985 | 38290.950724 |
| 5.235026 | 38290.965529 |
| 5.240067 | 38290.980194 |
| 5.245108 | 38290.994721 |
| 5.250149 | 38291.009110 |
| 5.255190 | 38291.023364 |
| 5.260231 | 38291.037484 |
| 5.265272 | 38291.051471 |
| 5.270313 | 38291.065326 |
| 5.275354 | 38291.079052 |
| 5.280395 | 38291.092649 |
| 5.285436 | 38291.106119 |
| 5.290477 | 38291.119464 |
| 5.295518 | 38291.132684 |
| 5.300559 | 38291.141967 |
| 5.305600 | 38291.154871 |
| 5.310641 | 38291.167671 |
| 5.315682 | 38291.180367 |
| 5.320724 | 38291.192960 |
| 5.325765 | 38291.205452 |
| 5.330806 | 38291.217843 |
| 5.335847 | 38291.230134 |
| 5.340888 | 38291.242326 |
| 5.345929 | 38291.254419 |
| 5.350970 | 38291.266414 |
| 5.356011 | 38291.278313 |
| 5.361052 | 38291.290116 |
| 5.366093 | 38291.301824 |
| 5.371134 | 38291.313437 |
| 5.376175 | 38291.324957 |
| 5.381216 | 38291.336384 |
| 5.386257 | 38291.347719 |
| 5.391298 | 38291.358963 |
| 5.396339 | 38291.370116 |
| 5.401380 | 38291.381180 |
| 5.406421 | 38291.392154 |
| 5.411462 | 38291.403041 |
| 5.416503 | 38291.413840 |
| 5.421544 | 38291.424552 |
| 5.426585 | 38291.435178 |
| 5.431626 | 38291.445720 |
| 5.436667 | 38291.456176 |
| 5.441708 | 38291.466549 |
| 5.446749 | 38291.476839 |
| 5.451790 | 38291.487046 |
| 5.456831 | 38291.497172 |
| 5.461872 | 38291.507217 |
| 5.466913 | 38291.517181 |
| 5.471954 | 38291.527066 |
| 5.476995 | 38291.536872 |
| 5.482036 | 38291.546600 |
| 5.487077 | 38291.556250 |
| 5.492118 | 38291.565824 |
| 5.497159 | 38291.575321 |
| 5.502200 | 38291.584742 |
| 5.507241 | 38291.594089 |
| 5.512282 | 38291.603362 |
| 5.517323 | 38291.612560 |
| 5.522364 | 38291.621686 |
| 5.527405 | 38291.630740 |
| 5.532446 | 38291.639722 |
| 5.537487 | 38291.648632 |
| 5.542528 | 38291.657473 |
| 5.547569 | 38291.666243 |
| 5.552610 | 38291.674944 |
| 5.557651 | 38291.683577 |
| 5.562692 | 38291.692141 |
| 5.567733 | 38291.700638 |
| 5.572774 | 38291.709069 |
| 5.577815 | 38291.717432 |
| 5.582856 | 38291.725731 |
| 5.587897 | 38291.733964 |
| 5.592938 | 38291.742132 |
| 5.597979 | 38291.750237 |
| 5.603020 | 38291.758278 |
| 5.608061 | 38291.766256 |
| 5.613102 | 38291.774172 |
| 5.618143 | 38291.782027 |
| 5.623184 | 38291.789820 |
| 5.628225 | 38291.797552 |
| 5.633266 | 38291.805225 |
| 5.638307 | 38291.812837 |
| 5.643348 | 38291.820391 |
| 5.648389 | 38291.827886 |
| 5.653430 | 38291.835323 |
| 5.658471 | 38291.842702 |
| 5.663512 | 38291.850025 |
| 5.668553 | 38291.857291 |
| 5.673594 | 38291.864501 |
| 5.678635 | 38291.871656 |
| 5.683676 | 38291.878755 |
| 5.688717 | 38291.885800 |
| 5.693758 | 38291.892791 |
| 5.698799 | 38291.899729 |
| 5.703840 | 38291.906613 |
| 5.708881 | 38291.913445 |
| 5.713922 | 38291.920225 |
| 5.718963 | 38291.926953 |
| 5.724004 | 38291.933630 |
| 5.729045 | 38291.940256 |
| 5.734086 | 38291.946832 |
| 5.739127 | 38291.953358 |
| 5.744169 | 38291.959835 |
| 5.749210 | 38291.966262 |
| 5.754251 | 38291.972642 |
| 5.759292 | 38291.978973 |
| 5.764333 | 38291.985256 |
| 5.769374 | 38291.991493 |
| 5.774415 | 38291.997682 |
| 5.779456 | 38292.003825 |
| 5.784497 | 38292.009923 |
| 5.789538 | 38292.015974 |
| 5.794579 | 38292.021981 |
| 5.799620 | 38292.027943 |
| 5.804661 | 38292.033860 |
| 5.809702 | 38292.039734 |
| 5.814743 | 38292.045564 |
| 5.819784 | 38292.051351 |
| 5.824825 | 38292.057096 |
| 5.829866 | 38292.062798 |
| 5.834907 | 38292.068458 |
| 5.839948 | 38292.074076 |
| 5.844989 | 38292.079653 |
| 5.850030 | 38292.085190 |
| 5.855071 | 38292.090685 |
| 5.860112 | 38292.096141 |
| 5.865153 | 38292.101557 |
| 5.870194 | 38292.106934 |
| 5.875235 | 38292.112271 |
| 5.880276 | 38292.117570 |
| 5.885317 | 38292.122831 |
| 5.890358 | 38292.128053 |
| 5.895399 | 38292.133238 |
| 5.900440 | 38292.138386 |
| 5.905481 | 38292.143497 |
| 5.910522 | 38292.148571 |
| 5.915563 | 38292.153608 |
| 5.920604 | 38292.158610 |
| 5.925645 | 38292.163576 |
| 5.930686 | 38292.168507 |
| 5.935727 | 38292.173403 |
| 5.940768 | 38292.178264 |
| 5.945809 | 38292.183090 |
| 5.950850 | 38292.187883 |
| 5.955891 | 38292.192642 |
| 5.960932 | 38292.197367 |
| 5.965973 | 38292.202059 |
| 5.971014 | 38292.206718 |
| 5.976055 | 38292.211345 |
| 5.981096 | 38292.215939 |
| 5.986137 | 38292.220502 |
| 5.991178 | 38292.225032 |
| 5.996219 | 38292.229532 |
| 6.001260 | 38292.234000 |
| 6.006301 | 38292.238437 |
| 6.011342 | 38292.242844 |
| 6.016383 | 38292.247220 |
| 6.021424 | 38292.251566 |
| 6.026465 | 38292.255883 |
| 6.031506 | 38292.260170 |
| 6.036547 | 38292.264427 |
| 6.041588 | 38292.268656 |
| 6.046629 | 38292.272856 |
| 6.051670 | 38292.277027 |
| 6.056711 | 38292.281171 |
| 6.061752 | 38292.285286 |
| 6.066793 | 38292.289374 |
| 6.071834 | 38292.293434 |
| 6.076875 | 38292.297466 |
| 6.081916 | 38292.301472 |
| 6.086957 | 38292.305451 |
| 6.091998 | 38292.309404 |
| 6.097039 | 38292.313330 |
| 6.102080 | 38292.317230 |
| 6.107121 | 38292.321104 |
| 6.112162 | 38292.324953 |
| 6.117203 | 38292.328776 |
| 6.122244 | 38292.332574 |
| 6.127285 | 38292.336347 |
| 6.132326 | 38292.340095 |
| 6.137367 | 38292.343819 |
| 6.142408 | 38292.347519 |
| 6.147449 | 38292.351194 |
| 6.152490 | 38292.354846 |
| 6.157531 | 38292.358474 |
| 6.162573 | 38292.362078 |
| 6.167614 | 38292.365659 |
| 6.172655 | 38292.369217 |
| 6.177696 | 38292.372752 |
| 6.182737 | 38292.376264 |
| 6.187778 | 38292.379754 |
| 6.192819 | 38292.383222 |
| 6.197860 | 38292.386667 |
| 6.202901 | 38292.390091 |
| 6.207942 | 38292.393493 |
| 6.212983 | 38292.396873 |
| 6.218024 | 38292.400232 |
| 6.223065 | 38292.403569 |
| 6.228106 | 38292.406886 |
| 6.233147 | 38292.410182 |
| 6.238188 | 38292.413457 |
| 6.243229 | 38292.416711 |
| 6.248270 | 38292.419946 |
| 6.253311 | 38292.423160 |
| 6.258352 | 38292.426354 |
| 6.263393 | 38292.429528 |
| 6.268434 | 38292.432683 |
| 6.273475 | 38292.435818 |
| 6.278516 | 38292.438934 |
| 6.283557 | 38292.442030 |
| 6.288598 | 38292.445108 |
| 6.293639 | 38292.448166 |
| 6.298680 | 38292.451206 |
| 6.303721 | 38292.454228 |
| 6.308762 | 38292.457230 |
| 6.313803 | 38292.460215 |
| 6.318844 | 38292.463181 |
| 6.323885 | 38292.466130 |
| 6.328926 | 38292.469061 |
| 6.333967 | 38292.471974 |
| 6.339008 | 38292.474869 |
| 6.344049 | 38292.477747 |
| 6.349090 | 38292.480607 |
| 6.354131 | 38292.483451 |
| 6.359172 | 38292.486277 |
| 6.364213 | 38292.489087 |
| 6.369254 | 38292.491880 |
| 6.374295 | 38292.494656 |
| 6.379336 | 38292.497415 |
| 6.384377 | 38292.500159 |
| 6.389418 | 38292.502886 |
| 6.394459 | 38292.505597 |
| 6.399500 | 38292.508292 |
| 6.404541 | 38292.510971 |
| 6.409582 | 38292.513634 |
| 6.414623 | 38292.516281 |
| 6.419664 | 38292.518913 |
| 6.424705 | 38292.521530 |
| 6.429746 | 38292.524131 |
| 6.434787 | 38292.526717 |
| 6.439828 | 38292.529288 |
| 6.444869 | 38292.531845 |
| 6.449910 | 38292.534386 |
| 6.454951 | 38292.536912 |
| 6.459992 | 38292.539424 |
| 6.465033 | 38292.541921 |
| 6.470074 | 38292.544404 |
| 6.475115 | 38292.546873 |
| 6.480156 | 38292.549327 |
| 6.485197 | 38292.551767 |
| 6.490238 | 38292.554193 |
| 6.495279 | 38292.556605 |
| 6.500320 | 38292.559004 |
| 6.505361 | 38292.561388 |
| 6.510402 | 38292.563759 |
| 6.515443 | 38292.566117 |
| 6.520484 | 38292.568461 |
| 6.525525 | 38292.570791 |
| 6.530566 | 38292.573109 |
| 6.535607 | 38292.575413 |
| 6.540648 | 38292.577704 |
| 6.545689 | 38292.579982 |
| 6.550730 | 38292.582247 |
| 6.555771 | 38292.584499 |
| 6.560812 | 38292.586739 |
| 6.565853 | 38292.588966 |
| 6.570894 | 38292.591180 |
| 6.575935 | 38292.593382 |
| 6.580976 | 38292.595571 |
| 6.586018 | 38292.597748 |
| 6.591059 | 38292.599913 |
| 6.596100 | 38292.602066 |
| 6.601141 | 38292.604206 |
| 6.606182 | 38292.606335 |
| 6.611223 | 38292.608451 |
| 6.616264 | 38292.610556 |
| 6.621305 | 38292.612649 |
| 6.626346 | 38292.614730 |
| 6.631387 | 38292.616800 |
| 6.636428 | 38292.618858 |
| 6.641469 | 38292.620904 |
| 6.646510 | 38292.622939 |
| 6.651551 | 38292.624963 |
| 6.656592 | 38292.626975 |
| 6.661633 | 38292.628976 |
| 6.666674 | 38292.630966 |
| 6.671715 | 38292.632945 |
| 6.676756 | 38292.634913 |
| 6.681797 | 38292.636870 |
| 6.686838 | 38292.638816 |
| 6.691879 | 38292.640751 |
| 6.696920 | 38292.642675 |
| 6.701961 | 38292.644589 |
| 6.707002 | 38292.646492 |
| 6.712043 | 38292.648384 |
| 6.717084 | 38292.650266 |
| 6.722125 | 38292.652138 |
| 6.727166 | 38292.653999 |
| 6.732207 | 38292.655849 |
| 6.737248 | 38292.657690 |
| 6.742289 | 38292.659520 |
| 6.747330 | 38292.661340 |
| 6.752371 | 38292.663150 |
| 6.757412 | 38292.664950 |
| 6.762453 | 38292.666740 |
| 6.767494 | 38292.668520 |
| 6.772535 | 38292.670290 |
| 6.777576 | 38292.672050 |
| 6.782617 | 38292.673801 |
| 6.787658 | 38292.675541 |
| 6.792699 | 38292.677272 |
| 6.797740 | 38292.678994 |
| 6.802781 | 38292.680706 |
| 6.807822 | 38292.682408 |
| 6.812863 | 38292.684101 |
| 6.817904 | 38292.685784 |
| 6.822945 | 38292.687458 |
| 6.827986 | 38292.689123 |
| 6.833027 | 38292.690779 |
| 6.838068 | 38292.692425 |
| 6.843109 | 38292.694062 |
| 6.848150 | 38292.695690 |
| 6.853191 | 38292.697309 |
| 6.858232 | 38292.698919 |
| 6.863273 | 38292.700520 |
| 6.868314 | 38292.702112 |
| 6.873355 | 38292.703695 |
| 6.878396 | 38292.705270 |
| 6.883437 | 38292.706835 |
| 6.888478 | 38292.708392 |
| 6.893519 | 38292.709940 |
| 6.898560 | 38292.711479 |
| 6.903601 | 38292.713010 |
| 6.908642 | 38292.714532 |
| 6.913683 | 38292.716046 |
| 6.918724 | 38292.717551 |
| 6.923765 | 38292.719048 |
| 6.928806 | 38292.720537 |
| 6.933847 | 38292.722017 |
| 6.938888 | 38292.723488 |
| 6.943929 | 38292.724952 |
| 6.948970 | 38292.726407 |
| 6.954011 | 38292.727854 |
| 6.959052 | 38292.729293 |
| 6.964093 | 38292.730724 |
| 6.969134 | 38292.732147 |
| 6.974175 | 38292.733562 |
| 6.979216 | 38292.734969 |
| 6.984257 | 38292.736368 |
| 6.989298 | 38292.737759 |
| 6.994339 | 38292.739142 |
| 6.999380 | 38292.740517 |
| 7.004422 | 38292.741885 |
| 7.009463 | 38292.743245 |
| 7.014504 | 38292.744597 |
| 7.019545 | 38292.745941 |
| 7.024586 | 38292.747278 |
| 7.029627 | 38292.748608 |
| 7.034668 | 38292.749930 |
| 7.039709 | 38292.751244 |
| 7.044750 | 38292.752551 |
| 7.049791 | 38292.753850 |
| 7.054832 | 38292.755142 |
| 7.059873 | 38292.756427 |
| 7.064914 | 38292.757705 |
| 7.069955 | 38292.758975 |
| 7.074996 | 38292.760238 |
| 7.080037 | 38292.761494 |
| 7.085078 | 38292.762742 |
| 7.090119 | 38292.763984 |
| 7.095160 | 38292.765218 |
| 7.100201 | 38292.766446 |
| 7.105242 | 38292.767666 |
| 7.110283 | 38292.768880 |
| 7.115324 | 38292.770087 |
| 7.120365 | 38292.771286 |
| 7.125406 | 38292.772479 |
| 7.130447 | 38292.773665 |
| 7.135488 | 38292.774845 |
| 7.140529 | 38292.776017 |
| 7.145570 | 38292.777183 |
| 7.150611 | 38292.778342 |
| 7.155652 | 38292.779495 |
| 7.160693 | 38292.780641 |
| 7.165734 | 38292.781781 |
| 7.170775 | 38292.782914 |
| 7.175816 | 38292.784041 |
| 7.180857 | 38292.785161 |
| 7.185898 | 38292.786275 |
| 7.190939 | 38292.787382 |
| 7.195980 | 38292.788483 |
| 7.201021 | 38292.789578 |
| 7.206062 | 38292.790667 |
| 7.211103 | 38292.791749 |
| 7.216144 | 38292.792825 |
| 7.221185 | 38292.793896 |
| 7.226226 | 38292.794960 |
| 7.231267 | 38292.796018 |
| 7.236308 | 38292.797070 |
| 7.241349 | 38292.798116 |
| 7.246390 | 38292.799156 |
| 7.251431 | 38292.800190 |
| 7.256472 | 38292.801219 |
| 7.261513 | 38292.802241 |
| 7.266554 | 38292.803258 |
| 7.271595 | 38292.804269 |
| 7.276636 | 38292.805275 |
| 7.281677 | 38292.806274 |
| 7.286718 | 38292.807269 |
| 7.291759 | 38292.808257 |
| 7.296800 | 38292.809240 |
| 7.301841 | 38292.810218 |
| 7.306882 | 38292.811190 |
| 7.311923 | 38292.812156 |
| 7.316964 | 38292.813117 |
| 7.322005 | 38292.814073 |
| 7.327046 | 38292.815024 |
| 7.332087 | 38292.815969 |
| 7.337128 | 38292.816909 |
| 7.342169 | 38292.817844 |
| 7.347210 | 38292.818774 |
| 7.352251 | 38292.819698 |
| 7.357292 | 38292.820618 |
| 7.362333 | 38292.821532 |
| 7.367374 | 38292.822442 |
| 7.372415 | 38292.823346 |
| 7.377456 | 38292.824246 |
| 7.382497 | 38292.825141 |
| 7.387538 | 38292.826030 |
| 7.392579 | 38292.826916 |
| 7.397620 | 38292.827796 |
| 7.402661 | 38292.828672 |
| 7.407702 | 38292.829543 |
| 7.412743 | 38292.830409 |
| 7.417784 | 38292.831453 |
| 7.422825 | 38292.832339 |
| 7.427867 | 38292.833220 |
| 7.432908 | 38292.834097 |
| 7.437949 | 38292.834970 |
| 7.442990 | 38292.835838 |
| 7.448031 | 38292.836703 |
| 7.453072 | 38292.837563 |
| 7.458113 | 38292.838419 |
| 7.463154 | 38292.839271 |
| 7.468195 | 38292.840119 |
| 7.473236 | 38292.840963 |
| 7.478277 | 38292.841803 |
| 7.483318 | 38292.842638 |
| 7.488359 | 38292.843470 |
| 7.493400 | 38292.844298 |
| 7.498441 | 38292.845122 |
| 7.503482 | 38292.845941 |
| 7.508523 | 38292.846757 |
| 7.513564 | 38292.847569 |
| 7.518605 | 38292.848377 |
| 7.523646 | 38292.849181 |
| 7.528687 | 38292.849981 |
| 7.533728 | 38292.850778 |
| 7.538769 | 38292.851570 |
| 7.543810 | 38292.852359 |
| 7.548851 | 38292.853144 |
| 7.553892 | 38292.853925 |
| 7.558933 | 38292.854703 |
| 7.563974 | 38292.855476 |
| 7.569015 | 38292.856246 |
| 7.574056 | 38292.857012 |
| 7.579097 | 38292.857775 |
| 7.584138 | 38292.858534 |
| 7.589179 | 38292.859289 |
| 7.594220 | 38292.860040 |
| 7.599261 | 38292.860788 |
| 7.604302 | 38292.861533 |
| 7.609343 | 38292.862273 |
| 7.614384 | 38292.863010 |
| 7.619425 | 38292.863744 |
| 7.624466 | 38292.864474 |
| 7.629507 | 38292.865201 |
| 7.634548 | 38292.865924 |
| 7.639589 | 38292.866643 |
| 7.644630 | 38292.867359 |
| 7.649671 | 38292.868072 |
| 7.654712 | 38292.868781 |
| 7.659753 | 38292.869487 |
| 7.664794 | 38292.870189 |
| 7.669835 | 38292.870888 |
| 7.674876 | 38292.871584 |
| 7.679917 | 38292.872276 |
| 7.684958 | 38292.872965 |
| 7.689999 | 38292.873650 |
| 7.695040 | 38292.874333 |
| 7.700081 | 38292.875012 |
| 7.705122 | 38292.875687 |
| 7.710163 | 38292.876360 |
| 7.715204 | 38292.877029 |
| 7.720245 | 38292.877695 |
| 7.725286 | 38292.878358 |
| 7.730327 | 38292.879017 |
| 7.735368 | 38292.879674 |
| 7.740409 | 38292.880327 |
| 7.745450 | 38292.880977 |
| 7.750491 | 38292.881624 |
| 7.755532 | 38292.882268 |
| 7.760573 | 38292.882909 |
| 7.765614 | 38292.883547 |
| 7.770655 | 38292.884181 |
| 7.775696 | 38292.884813 |
| 7.780737 | 38292.885441 |
| 7.785778 | 38292.886067 |
| 7.790819 | 38292.886689 |
| 7.795860 | 38292.887309 |
| 7.800901 | 38292.887925 |
| 7.805942 | 38292.888539 |
| 7.810983 | 38292.889150 |
| 7.816024 | 38292.889758 |
| 7.821065 | 38292.890362 |
| 7.826106 | 38292.890964 |
| 7.831147 | 38292.891563 |
| 7.836188 | 38292.892159 |
| 7.841229 | 38292.892753 |
| 7.846271 | 38292.893343 |
| 7.851312 | 38292.893931 |
| 7.856353 | 38292.894516 |
| 7.861394 | 38292.895098 |
| 7.866435 | 38292.895677 |
| 7.871476 | 38292.896254 |
| 7.876517 | 38292.896827 |
| 7.881558 | 38292.897398 |
| 7.886599 | 38292.897967 |
| 7.891640 | 38292.898532 |
| 7.896681 | 38292.899095 |
| 7.901722 | 38292.899655 |
| 7.906763 | 38292.900213 |
| 7.911804 | 38292.900768 |
| 7.916845 | 38292.901320 |
| 7.921886 | 38292.901870 |
| 7.926927 | 38292.902417 |
| 7.931968 | 38292.902961 |
| 7.937009 | 38292.903503 |
| 7.942050 | 38292.904042 |
| 7.947091 | 38292.904579 |
| 7.952132 | 38292.905113 |
| 7.957173 | 38292.905645 |
| 7.962214 | 38292.906174 |
| 7.967255 | 38292.906701 |
| 7.972296 | 38292.907225 |
| 7.977337 | 38292.907747 |
| 7.982378 | 38292.908266 |
| 7.987419 | 38292.908783 |
| 7.992460 | 38292.909297 |
| 7.997501 | 38292.909809 |
| 8.002542 | 38292.910319 |
| 8.007583 | 38292.910826 |
| 8.012624 | 38292.911330 |
| 8.017665 | 38292.911833 |
| 8.022706 | 38292.912333 |
| 8.027747 | 38292.912831 |
| 8.032788 | 38292.913326 |
| 8.037829 | 38292.913819 |
| 8.042870 | 38292.914310 |
| 8.047911 | 38292.914798 |
| 8.052952 | 38292.915284 |
| 8.057993 | 38292.915768 |
| 8.063034 | 38292.916250 |
| 8.068075 | 38292.916729 |
| 8.073116 | 38292.917206 |
| 8.078157 | 38292.917681 |
| 8.083198 | 38292.918154 |
| 8.088239 | 38292.918625 |
| 8.093280 | 38292.919093 |
| 8.098321 | 38292.919559 |
| 8.103362 | 38292.920023 |
| 8.108403 | 38292.920485 |
| 8.113444 | 38292.920945 |
| 8.118485 | 38292.921403 |
| 8.123526 | 38292.921858 |
| 8.128567 | 38292.922312 |
| 8.133608 | 38292.922763 |
| 8.138649 | 38292.923213 |
| 8.143690 | 38292.923660 |
| 8.148731 | 38292.924105 |
| 8.153772 | 38292.924548 |
| 8.158813 | 38292.924989 |
| 8.163854 | 38292.925428 |
| 8.168895 | 38292.925865 |
| 8.173936 | 38292.926301 |
| 8.178977 | 38292.926734 |
| 8.184018 | 38292.927165 |
| 8.189059 | 38292.927594 |
| 8.194100 | 38292.928021 |
| 8.199141 | 38292.928447 |
| 8.204182 | 38292.928870 |
| 8.209223 | 38292.929292 |
| 8.214264 | 38292.929711 |
| 8.219305 | 38292.930129 |
| 8.224346 | 38292.930545 |
| 8.229387 | 38292.930959 |
| 8.234428 | 38292.931371 |
| 8.239469 | 38292.931781 |
| 8.244510 | 38292.932189 |
| 8.249551 | 38292.932596 |
| 8.254592 | 38292.933001 |
| 8.259633 | 38292.933404 |
| 8.264674 | 38292.933805 |
| 8.269716 | 38292.934204 |
| 8.274757 | 38292.934602 |
| 8.279798 | 38292.934997 |
| 8.284839 | 38292.935391 |
| 8.289880 | 38292.935784 |
| 8.294921 | 38292.936174 |
| 8.299962 | 38292.936563 |
| 8.305003 | 38292.936950 |
| 8.310044 | 38292.937336 |
| 8.315085 | 38292.937719 |
| 8.320126 | 38292.938101 |
| 8.325167 | 38292.938482 |
| 8.330208 | 38292.938860 |
| 8.335249 | 38292.939237 |
| 8.340290 | 38292.939613 |
| 8.345331 | 38292.939987 |
| 8.350372 | 38292.940359 |
| 8.355413 | 38292.940729 |
| 8.360454 | 38292.941098 |
| 8.365495 | 38292.941465 |
| 8.370536 | 38292.941831 |
| 8.375577 | 38292.942195 |
| 8.380618 | 38292.942558 |
| 8.385659 | 38292.942918 |
| 8.390700 | 38292.943278 |
| 8.395741 | 38292.943636 |
| 8.400782 | 38292.943992 |
| 8.405823 | 38292.944347 |
| 8.410864 | 38292.944700 |
| 8.415905 | 38292.945052 |
| 8.420946 | 38292.945402 |
| 8.425987 | 38292.945751 |
| 8.431028 | 38292.946098 |
| 8.436069 | 38292.946444 |
| 8.441110 | 38292.946788 |
| 8.446151 | 38292.947131 |
| 8.451192 | 38292.947473 |
| 8.456233 | 38292.947813 |
| 8.461274 | 38292.948151 |
| 8.466315 | 38292.948488 |
| 8.471356 | 38292.948824 |
| 8.476397 | 38292.949158 |
| 8.481438 | 38292.949491 |
| 8.486479 | 38292.949823 |
| 8.491520 | 38292.950153 |
| 8.496561 | 38292.950481 |
| 8.501602 | 38292.950809 |
| 8.506643 | 38292.951135 |
| 8.511684 | 38292.951459 |
| 8.516725 | 38292.951783 |
| 8.521766 | 38292.952105 |
| 8.526807 | 38292.952425 |
| 8.531848 | 38292.952744 |
| 8.536889 | 38292.953062 |
| 8.541930 | 38292.953379 |
| 8.546971 | 38292.953694 |
| 8.552012 | 38292.954009 |
| 8.557053 | 38292.954321 |
| 8.562094 | 38292.954633 |
| 8.567135 | 38292.954943 |
| 8.572176 | 38292.955252 |
| 8.577217 | 38292.955560 |
| 8.582258 | 38292.955866 |
| 8.587299 | 38292.956171 |
| 8.592340 | 38292.956475 |
| 8.597381 | 38292.956778 |
| 8.602422 | 38292.957080 |
| 8.607463 | 38292.957380 |
| 8.612504 | 38292.957679 |
| 8.617545 | 38292.957977 |
| 8.622586 | 38292.958274 |
| 8.627627 | 38292.958569 |
| 8.632668 | 38292.958863 |
| 8.637709 | 38292.959156 |
| 8.642750 | 38292.959448 |
| 8.647791 | 38292.959739 |
| 8.652832 | 38292.960029 |
| 8.657873 | 38292.960317 |
| 8.662914 | 38292.960605 |
| 8.667955 | 38292.960891 |
| 8.672996 | 38292.961176 |
| 8.678037 | 38292.961460 |
| 8.683078 | 38292.961743 |
| 8.688119 | 38292.962024 |
| 8.693161 | 38292.962305 |
| 8.698202 | 38292.962584 |
| 8.703243 | 38292.962863 |
| 8.708284 | 38292.963140 |
| 8.713325 | 38292.963416 |
| 8.718366 | 38292.963691 |
| 8.723407 | 38292.963966 |
| 8.728448 | 38292.964238 |
| 8.733489 | 38292.964510 |
| 8.738530 | 38292.964781 |
| 8.743571 | 38292.965051 |
| 8.748612 | 38292.965320 |
| 8.753653 | 38292.965587 |
| 8.758694 | 38292.965854 |
| 8.763735 | 38292.966120 |
| 8.768776 | 38292.966384 |
| 8.773817 | 38292.966648 |
| 8.778858 | 38292.966910 |
| 8.783899 | 38292.967172 |
| 8.788940 | 38292.967433 |
| 8.793981 | 38292.967692 |
| 8.799022 | 38292.967951 |
| 8.804063 | 38292.968208 |
| 8.809104 | 38292.968465 |
| 8.814145 | 38292.968720 |
| 8.819186 | 38292.968975 |
| 8.824227 | 38292.969228 |
| 8.829268 | 38292.969481 |
| 8.834309 | 38292.969733 |
| 8.839350 | 38292.969983 |
| 8.844391 | 38292.970233 |
| 8.849432 | 38292.970482 |
| 8.854473 | 38292.970730 |
| 8.859514 | 38292.970977 |
| 8.864555 | 38292.971223 |
| 8.869596 | 38292.971468 |
| 8.874637 | 38292.971712 |
| 8.879678 | 38292.971955 |
| 8.884719 | 38292.972197 |
| 8.889760 | 38292.972438 |
| 8.894801 | 38292.972679 |
| 8.899842 | 38292.972918 |
| 8.904883 | 38292.973157 |
| 8.909924 | 38292.973394 |
| 8.914965 | 38292.973631 |
| 8.920006 | 38292.973867 |
| 8.925047 | 38292.974102 |
| 8.930088 | 38292.974336 |
| 8.935129 | 38292.974569 |
| 8.940170 | 38292.974801 |
| 8.945211 | 38292.975032 |
| 8.950252 | 38292.975263 |
| 8.955293 | 38292.975492 |
| 8.960334 | 38292.975721 |
| 8.965375 | 38292.975949 |
| 8.970416 | 38292.976176 |
| 8.975457 | 38292.976402 |
| 8.980498 | 38292.976627 |
| 8.985539 | 38292.976852 |
| 8.990580 | 38292.977075 |
| 8.995621 | 38292.977298 |
| 9.000662 | 38292.977520 |
| 9.005703 | 38292.977716 |
| 9.010744 | 38292.977932 |
| 9.015785 | 38292.978147 |
| 9.020826 | 38292.978361 |
| 9.025867 | 38292.978575 |
| 9.030908 | 38292.978788 |
| 9.035949 | 38292.979000 |
| 9.040990 | 38292.979211 |
| 9.046031 | 38292.979421 |
| 9.051072 | 38292.979631 |
| 9.056113 | 38292.979839 |
| 9.061154 | 38292.980047 |
| 9.066195 | 38292.980255 |
| 9.071236 | 38292.980461 |
| 9.076277 | 38292.980667 |
| 9.081318 | 38292.980872 |
| 9.086359 | 38292.981076 |
| 9.091400 | 38292.981280 |
| 9.096441 | 38292.981482 |
| 9.101482 | 38292.981684 |
| 9.106523 | 38292.981885 |
| 9.111565 | 38292.982086 |
| 9.116606 | 38292.982285 |
| 9.121647 | 38292.982484 |
| 9.126688 | 38292.982683 |
| 9.131729 | 38292.982880 |
| 9.136770 | 38292.983077 |
| 9.141811 | 38292.983273 |
| 9.146852 | 38292.983468 |
| 9.151893 | 38292.983663 |
| 9.156934 | 38292.983856 |
| 9.161975 | 38292.984049 |
| 9.167016 | 38292.984242 |
| 9.172057 | 38292.984433 |
| 9.177098 | 38292.984624 |
| 9.182139 | 38292.984814 |
| 9.187180 | 38292.985004 |
| 9.192221 | 38292.985192 |
| 9.197262 | 38292.985380 |
| 9.202303 | 38292.985568 |
| 9.207344 | 38292.985754 |
| 9.212385 | 38292.985940 |
| 9.217426 | 38292.986125 |
| 9.222467 | 38292.986310 |
| 9.227508 | 38292.986493 |
| 9.232549 | 38292.986676 |
| 9.237590 | 38292.986859 |
| 9.242631 | 38292.987040 |
| 9.247672 | 38292.987221 |
| 9.252713 | 38292.987402 |
| 9.257754 | 38292.987581 |
| 9.262795 | 38292.987760 |
| 9.267836 | 38292.987938 |
| 9.272877 | 38292.988116 |
| 9.277918 | 38292.988293 |
| 9.282959 | 38292.988469 |
| 9.288000 | 38292.988644 |
| 9.293041 | 38292.988819 |
| 9.298082 | 38292.988993 |
| 9.303123 | 38292.989167 |
| 9.308164 | 38292.989339 |
| 9.313205 | 38292.989512 |
| 9.318246 | 38292.989683 |
| 9.323287 | 38292.989854 |
| 9.328328 | 38292.990024 |
| 9.333369 | 38292.990193 |
| 9.338410 | 38292.990362 |
| 9.343451 | 38292.990530 |
| 9.348492 | 38292.990698 |
| 9.353533 | 38292.990865 |
| 9.358574 | 38292.991031 |
| 9.363615 | 38292.991197 |
| 9.368656 | 38292.991361 |
| 9.373697 | 38292.991526 |
| 9.378738 | 38292.991689 |
| 9.383779 | 38292.991852 |
| 9.388820 | 38292.992015 |
| 9.393861 | 38292.992176 |
| 9.398902 | 38292.992338 |
| 9.403943 | 38292.992498 |
| 9.408984 | 38292.992658 |
| 9.414025 | 38292.992817 |
| 9.419066 | 38292.992976 |
| 9.424107 | 38292.993134 |
| 9.429148 | 38292.993291 |
| 9.434189 | 38292.993448 |
| 9.439230 | 38292.993604 |
| 9.444271 | 38292.993759 |
| 9.449312 | 38292.993914 |
| 9.454353 | 38292.994069 |
| 9.459394 | 38292.994222 |
| 9.464435 | 38292.994375 |
| 9.469476 | 38292.994528 |
| 9.474517 | 38292.994680 |
| 9.479558 | 38292.994831 |
| 9.484599 | 38292.994982 |
| 9.489640 | 38292.995132 |
| 9.494681 | 38292.995281 |
| 9.499722 | 38292.995430 |
| 9.504763 | 38292.995578 |
| 9.509804 | 38292.995726 |
| 9.514845 | 38292.995873 |
| 9.519886 | 38292.996020 |
| 9.524927 | 38292.996166 |
| 9.529968 | 38292.996311 |
| 9.535010 | 38292.996456 |
| 9.540051 | 38292.996600 |
| 9.545092 | 38292.996744 |
| 9.550133 | 38292.996887 |
| 9.555174 | 38292.997030 |
| 9.560215 | 38292.997172 |
| 9.565256 | 38292.997313 |
| 9.570297 | 38292.997454 |
| 9.575338 | 38292.997594 |
| 9.580379 | 38292.997734 |
| 9.585420 | 38292.997873 |
| 9.590461 | 38292.998012 |
| 9.595502 | 38292.998150 |
| 9.600543 | 38292.998287 |
| 9.605584 | 38292.998424 |
| 9.610625 | 38292.998561 |
| 9.615666 | 38292.998697 |
| 9.620707 | 38292.998832 |
| 9.625748 | 38292.998967 |
| 9.630789 | 38292.999101 |
| 9.635830 | 38292.999235 |
| 9.640871 | 38292.999368 |
| 9.645912 | 38292.999501 |
| 9.650953 | 38292.999633 |
| 9.655994 | 38292.999765 |
| 9.661035 | 38292.999896 |
| 9.666076 | 38293.000026 |
| 9.671117 | 38293.000156 |
| 9.676158 | 38293.000286 |
| 9.681199 | 38293.000415 |
| 9.686240 | 38293.000544 |
| 9.691281 | 38293.000672 |
| 9.696322 | 38293.000799 |
| 9.701363 | 38293.000926 |
| 9.706404 | 38293.001053 |
| 9.711445 | 38293.001179 |
| 9.716486 | 38293.001304 |
| 9.721527 | 38293.001429 |
| 9.726568 | 38293.001554 |
| 9.731609 | 38293.001678 |
| 9.736650 | 38293.001801 |
| 9.741691 | 38293.001924 |
| 9.746732 | 38293.002047 |
| 9.751773 | 38293.002169 |
| 9.756814 | 38293.002290 |
| 9.761855 | 38293.002411 |
| 9.766896 | 38293.002532 |
| 9.771937 | 38293.002652 |
| 9.776978 | 38293.002772 |
| 9.782019 | 38293.002891 |
| 9.787060 | 38293.003010 |
| 9.792101 | 38293.003128 |
| 9.797142 | 38293.003246 |
| 9.802183 | 38293.003363 |
| 9.807224 | 38293.003480 |
| 9.812265 | 38293.003596 |
| 9.817306 | 38293.003712 |
| 9.822347 | 38293.003828 |
| 9.827388 | 38293.003943 |
| 9.832429 | 38293.004057 |
| 9.837470 | 38293.004171 |
| 9.842511 | 38293.004285 |
| 9.847552 | 38293.004398 |
| 9.852593 | 38293.004511 |
| 9.857634 | 38293.004623 |
| 9.862675 | 38293.004735 |
| 9.867716 | 38293.004847 |
| 9.872757 | 38293.004958 |
| 9.877798 | 38293.005068 |
| 9.882839 | 38293.005178 |
| 9.887880 | 38293.005288 |
| 9.892921 | 38293.005397 |
| 9.897962 | 38293.005506 |
| 9.903003 | 38293.005615 |
| 9.908044 | 38293.005723 |
| 9.913085 | 38293.005830 |
| 9.918126 | 38293.005938 |
| 9.923167 | 38293.006044 |
| 9.928208 | 38293.006151 |
| 9.933249 | 38293.006257 |
| 9.938290 | 38293.006362 |
| 9.943331 | 38293.006467 |
| 9.948372 | 38293.006572 |
| 9.953414 | 38293.006676 |
| 9.958455 | 38293.006780 |
| 9.963496 | 38293.006884 |
| 9.968537 | 38293.006987 |
| 9.973578 | 38293.007090 |
| 9.978619 | 38293.007192 |
| 9.983660 | 38293.007294 |
| 9.988701 | 38293.007396 |
| 9.993742 | 38293.007497 |
| 9.998783 | 38293.007598 |
| 10.003824 | 38293.007698 |
| 10.008865 | 38293.007798 |
| 10.013906 | 38293.007898 |
| 10.018947 | 38293.007997 |
| 10.023988 | 38293.008096 |
| 10.029029 | 38293.008194 |
| 10.034070 | 38293.008293 |
| 10.039111 | 38293.008390 |
| 10.044152 | 38293.008488 |
| 10.049193 | 38293.008585 |
| 10.054234 | 38293.008682 |
| 10.059275 | 38293.008778 |
| 10.064316 | 38293.008874 |
| 10.069357 | 38293.008970 |
| 10.074398 | 38293.009065 |
| 10.079439 | 38293.009160 |
| 10.084480 | 38293.009254 |
| 10.089521 | 38293.009349 |
| 10.094562 | 38293.009443 |
| 10.099603 | 38293.009536 |
| 10.104644 | 38293.009629 |
| 10.109685 | 38293.009722 |
| 10.114726 | 38293.009815 |
| 10.119767 | 38293.009907 |
| 10.124808 | 38293.009999 |
| 10.129849 | 38293.010091 |
| 10.134890 | 38293.010182 |
| 10.139931 | 38293.010273 |
| 10.144972 | 38293.010363 |
| 10.150013 | 38293.010453 |
| 10.155054 | 38293.010543 |
| 10.160095 | 38293.010633 |
| 10.165136 | 38293.010722 |
| 10.170177 | 38293.010811 |
| 10.175218 | 38293.010900 |
| 10.180259 | 38293.010988 |
| 10.185300 | 38293.011076 |
| 10.190341 | 38293.011164 |
| 10.195382 | 38293.011252 |
| 10.200423 | 38293.011339 |
| 10.205464 | 38293.011426 |
| 10.210505 | 38293.011512 |
| 10.215546 | 38293.011599 |
| 10.220587 | 38293.011685 |
| 10.225628 | 38293.011770 |
| 10.230669 | 38293.011856 |
| 10.235710 | 38293.011941 |
| 10.240751 | 38293.012026 |
| 10.245792 | 38293.012110 |
| 10.250833 | 38293.012194 |
| 10.255874 | 38293.012278 |
| 10.260915 | 38293.012362 |
| 10.265956 | 38293.012446 |
| 10.270997 | 38293.012529 |
| 10.276038 | 38293.012612 |
| 10.281079 | 38293.012694 |
| 10.286120 | 38293.012777 |
| 10.291161 | 38293.012859 |
| 10.296202 | 38293.012941 |
| 10.301243 | 38293.013022 |
| 10.306284 | 38293.013104 |
| 10.311325 | 38293.013185 |
| 10.316366 | 38293.013266 |
| 10.321407 | 38293.013346 |
| 10.326448 | 38293.013426 |
| 10.331489 | 38293.013507 |
| 10.336530 | 38293.013586 |
| 10.341571 | 38293.013666 |
| 10.346612 | 38293.013745 |
| 10.351653 | 38293.013824 |
| 10.356694 | 38293.013903 |
| 10.361735 | 38293.013982 |
| 10.366776 | 38293.014060 |
| 10.371817 | 38293.014139 |
| 10.376859 | 38293.014217 |
| 10.381900 | 38293.014294 |
| 10.386941 | 38293.014372 |
| 10.391982 | 38293.014449 |
| 10.397023 | 38293.014526 |
| 10.402064 | 38293.014603 |
| 10.407105 | 38293.014680 |
| 10.412146 | 38293.014756 |
| 10.417187 | 38293.014832 |
| 10.422228 | 38293.014908 |
| 10.427269 | 38293.014984 |
| 10.432310 | 38293.015060 |
| 10.437351 | 38293.015135 |
| 10.442392 | 38293.015210 |
| 10.447433 | 38293.015285 |
| 10.452474 | 38293.015360 |
| 10.457515 | 38293.015435 |
| 10.462556 | 38293.015509 |
| 10.467597 | 38293.015583 |
| 10.472638 | 38293.015657 |
| 10.477679 | 38293.015731 |
| 10.482720 | 38293.015805 |
| 10.487761 | 38293.015878 |
| 10.492802 | 38293.015952 |
| 10.497843 | 38293.016025 |
| 10.502884 | 38293.016098 |
| 10.507925 | 38293.016171 |
| 10.512966 | 38293.016243 |
| 10.518007 | 38293.016316 |
| 10.523048 | 38293.016388 |
| 10.528089 | 38293.016460 |
| 10.533130 | 38293.016532 |
| 10.538171 | 38293.016604 |
| 10.543212 | 38293.016675 |
| 10.548253 | 38293.016747 |
| 10.553294 | 38293.016818 |
| 10.558335 | 38293.016889 |
| 10.563376 | 38293.016960 |
| 10.568417 | 38293.017031 |
| 10.573458 | 38293.017102 |
| 10.578499 | 38293.017173 |
| 10.583540 | 38293.017243 |

Table S2. Theoretical values of $r_e$ and $<r>$ internuclear distances for $^nH_2$, $n$ = 1–3.

| Isotopologue | $^1H_2$ | $^2H_2$ | $^3H_2$ |
|---|---|---|---|
| Equilibrium internuclear distances $r_e$, Å | | | |
| Kołos et al. (1964)<br>10.1063/1.1725796 | 0.74141 | – | – |
| Sims et al. (2006)<br>10.1063/1.2173250 | 0.74144 | – | – |
| Pachucki (2010)<br>10.1103/PhysRevA.82.032509<br>Pachucki et al. (2014)<br>10.1063/1.4902981 | 0.74144 | 0.74144* | 0.74144* |
| Kurokawa et al. (2018)<br>10.1039/c8cp05949g | 0.74144 | – | – |
| (Ro)vibrationally averaged over the ground state internuclear distances $<r>$, Å | | | |
| Kołos et al. (1964)<br>10.1063/1.1725797 | 0.76661 | 0.75912 | 0.75584 |
| Bubin (2003)<br>10.1063/1.1537719 | 0.76664 | – | – |
| Kdziera et al. (2006)<br>10.1063/1.2209691 | 0.76664 | – | – |
| Alexander et al. (2008)<br>10.1063/1.2978172 | 0.76667 | 0.75916 | 0.75593 |
| Bubin (2011)<br>10.1063/1.3625955 | – | 0.75914 | – |
| Bubin (2014)<br>10.1063/1.4870935 | – | – | 0.75585 |
| Kirnosov (2015)<br>10.1016/j.cplett.2015.10.017 | – | – | 0.75588 |
| Kurokawa et al. (2018)<br>10.1039/c8cp05949g | 0.76646 | – | – |
| This work | 0.76682 | 0.75941 | 0.75616 |

*Due to the Born-Oppenheimer approximation used.

Table S3. Calculated vibrational energies ($G(\upsilon)$) for $^1H_2$–$^7H_2$ isotopologues in the Pachucki's U($r$) potential.

| Vib. state ($\upsilon$) | $G(\upsilon)$, cm$^{-1}$ | | | | | | |
|---|---|---|---|---|---|---|---|
| | $^1H_2$ | $^2H_2$ | $^3H_2$ | $^4H_2$ | $^5H_2$ | $^6H_2$ | $^7H_2$ |
| 0 | 2180.904 | 1547.501 | 1266.134 | 1096.700 | 981.256 | 895.880 | 829.819 |
| 1 | 6345.216 | 4542.890 | 3732.151 | 3240.308 | 2903.854 | 2654.531 | 2460.735 |
| 2 | 10273.812 | 7419.105 | 6118.060 | 5323.781 | 4778.387 | 4373.017 | 4056.974 |
| 3 | 13971.301 | 10178.121 | 8424.958 | 7347.996 | 6605.074 | 6051.559 | 5619.194 |
| 4 | 17441.195 | 12821.693 | 10654.161 | 9313.391 | 8384.794 | 7690.815 | 7147.615 |
| 5 | 20684.152 | 15350.699 | 12806.330 | 11220.845 | 10117.985 | 9291.224 | 8642.676 |
| 6 | 23698.856 | 17766.237 | 14882.340 | 13070.797 | 11804.867 | 10852.786 | 10104.158 |
| 7 | 26481.574 | 20068.306 | 16882.193 | 14863.685 | 13445.879 | 12375.940 | 11532.938 |
| 8 | 29024.627 | 22256.468 | 18806.327 | 16599.729 | 15041.240 | 13861.125 | 12928.797 |
| 9 | 31317.040 | 24329.626 | 20654.743 | 18278.710 | 16590.950 | 15308.340 | 14291.953 |
| 10 | 33342.132 | 26286.023 | 22426.781 | 19901.067 | 18095.449 | 16717.587 | 15622.848 |
| 11 | 35077.079 | 28123.025 | 24122.003 | 21466.579 | 19554.736 | 18089.084 | 16921.040 |
| 12 | 36489.398 | 29837.122 | 25739.531 | 22974.809 | 20968.372 | 19422.831 | 18186.970 |
| 13 | 37535.195 | 31423.046 | 27278.048 | 24425.317 | 22336.138 | 20718.610 | 19420.417 |
| 14 | 38152.357 | 32875.090 | 28735.359 | 25817.444 | 23658.253 | 21976.419 | 20621.382 |
| 15 | | 34185.354 | 30109.710 | 27150.094 | 24933.839 | 23196.039 | 21789.865 |
| 16 | | 35343.521 | 31397.806 | 28422.389 | 26162.458 | 24377.032 | 22925.647 |
| 17 | | 36337.302 | 32596.796 | 29632.572 | 27343.232 | 25519.178 | 24028.507 |
| 18 | | 37150.895 | 33702.070 | 30779.327 | 28475.721 | 26622.038 | 25098.006 |
| 19 | | 37764.546 | 34708.581 | 31860.459 | 29558.609 | 27684.734 | 26133.927 |
| 20 | | 38153.016 | 35610.402 | 32873.334 | 30590.579 | 28707.047 | 27136.048 |
| 21 | | 38291.724 | 36399.853 | 33815.319 | 31570.533 | 29687.660 | 28103.492 |
| 22 | | | 37068.372 | 34683.122 | 32496.716 | 30625.914 | 29035.820 |
| 23 | | | 37604.988 | 35472.353 | 33366.933 | 31520.712 | 29932.155 |
| 24 | | | 37996.750 | 36178.622 | 34178.989 | 32370.737 | 30792.276 |
| 25 | | | 38228.515 | 36796.443 | 34930.470 | 33174.453 | 31614.647 |
| 26 | | | | 37319.012 | 35618.303 | 33930.324 | 32398.611 |
| 27 | | | | 37738.867 | 36238.758 | 34636.374 | 33142.849 |
| 28 | | | | 38046.790 | 36788.103 | 35290.189 | 33846.046 |
| 29 | | | | 38232.905 | 37261.290 | 35889.794 | 34506.884 |
| 30 | | | | 38292.602 | 37653.272 | 36431.896 | 35123.388 |
| 31 | | | | | 37957.683 | 36913.423 | 35693.803 |
| 32 | | | | | 38167.282 | 37330.644 | 36215.933 |
| 33 | | | | | 38275.044 | 37679.170 | 36687.364 |
| 34 | | | | | | 37954.391 | 37105.244 |
| 35 | | | | | | 38150.602 | 37466.280 |
| 36 | | | | | | 38262.753 | 37767.180 |
| 37 | | | | | | | 38003.993 |
| 38 | | | | | | | 38172.110 |
| 39 | | | | | | | 38267.362 |

Table S4. Calculated vibrational energies ($G(\upsilon)$) for $^1H_2$–$^7H_2$ isotopologues in the M1($r$) potential.

| Vib. state ($\upsilon$) | $G(\upsilon)$, $cm^{-1}$ | | | | | | |
|---|---|---|---|---|---|---|---|
| | $^1H_2$ | $^2H_2$ | $^3H_2$ | $^4H_2$ | $^5H_2$ | $^6H_2$ | $^7H_2$ |
| 0 | 2170.549 | 1542.385 | 1263.049 | 1094.355 | 979.323 | 894.387 | 828.462 |
| 1 | 6334.861 | 4537.775 | 3729.066 | 3237.964 | 2901.921 | 2653.037 | 2459.378 |
| 2 | 10263.457 | 7413.990 | 6114.975 | 5321.437 | 4776.454 | 4371.523 | 4055.617 |
| 3 | 13956.337 | 10171.031 | 8420.775 | 7344.773 | 6602.922 | 6049.846 | 5617.179 |
| 4 | 17413.501 | 12808.896 | 10646.467 | 9307.974 | 8381.325 | 7688.004 | 7144.064 |
| 5 | 20634.950 | 15327.587 | 12792.051 | 11211.039 | 10111.663 | 9285.999 | 8636.272 |
| 6 | 23620.683 | 17727.103 | 14857.527 | 13053.967 | 11793.936 | 10843.830 | 10093.803 |
| 7 | 26370.700 | 20007.445 | 16842.895 | 14836.759 | 13428.144 | 12361.497 | 11516.657 |
| 8 | 28885.001 | 22168.611 | 18748.154 | 16559.416 | 15014.287 | 13839.000 | 12904.834 |
| 9 | 31163.587 | 24210.603 | 20573.305 | 18221.936 | 16552.365 | 15276.340 | 14258.334 |
| 10 | 33206.457 | 26133.420 | 22318.348 | 19824.320 | 18042.378 | 16673.515 | 15577.157 |
| 11 | 35013.611 | 27937.063 | 23983.282 | 21366.569 | 19484.327 | 18030.527 | 16861.303 |
| 12 | 36585.049 | 29621.531 | 25568.109 | 22848.681 | 20878.210 | 19347.375 | 18110.772 |
| 13 | 37920.772 | 31186.824 | 27072.827 | 24270.657 | 22224.029 | 20624.059 | 19325.564 |
| 14 | 39020.779 | 32632.942 | 28497.437 | 25632.497 | 23521.782 | 21860.579 | 20505.679 |
| 15 | 39885.070 | 33959.886 | 29841.938 | 26934.201 | 24771.471 | 23056.935 | 21651.118 |
| 16 | 40513.645 | 35167.655 | 31106.332 | 28175.769 | 25973.094 | 24213.127 | 22761.879 |
| 17 | 40906.505 | 36256.249 | 32290.617 | 29357.201 | 27126.653 | 25329.156 | 23837.963 |
| 18 | 41063.649 | 37225.668 | 33394.794 | 30478.497 | 28232.147 | 26405.020 | 24879.370 |
| 19 | | 38075.913 | 34418.862 | 31539.657 | 29289.575 | 27440.721 | 25886.100 |
| 20 | | 38806.983 | 35362.823 | 32540.681 | 30298.939 | 28436.258 | 26858.153 |
| 21 | | 39418.878 | 36226.675 | 33481.568 | 31260.238 | 29391.631 | 27795.529 |
| 22 | | 39911.599 | 37010.419 | 34362.320 | 32173.472 | 30306.840 | 28698.228 |
| 23 | | 40285.145 | 37714.054 | 35182.936 | 33038.641 | 31181.886 | 29566.251 |
| 24 | | 40539.516 | 38337.582 | 35943.415 | 33855.745 | 32016.767 | 30399.596 |
| 25 | | 40674.712 | 38881.001 | 36643.759 | 34624.784 | 32811.485 | 31198.264 |
| 26 | | 40690.734 | 39344.312 | 37283.966 | 35345.758 | 33566.039 | 31962.255 |
| 27 | | | 39727.515 | 37864.038 | 36018.668 | 34280.429 | 32691.569 |
| 28 | | | 40030.609 | 38383.973 | 36643.512 | 34954.655 | 33386.207 |
| 29 | | | 40253.595 | 38843.773 | 37220.291 | 35588.717 | 34046.167 |
| 30 | | | 40396.473 | 39243.436 | 37749.006 | 36182.615 | 34671.450 |
| 31 | | | 40459.243 | 39582.963 | 38229.655 | 36736.350 | 35262.056 |
| 32 | | | | 39862.354 | 38662.240 | 37249.920 | 35817.985 |
| 33 | | | | 40081.609 | 39046.759 | 37723.327 | 36339.238 |
| 34 | | | | 40240.729 | 39383.214 | 38156.570 | 36825.813 |
| 35 | | | | 40339.712 | 39671.603 | 38549.649 | 37277.711 |
| 36 | | | | 40378.559 | 39911.928 | 38902.564 | 37694.933 |
| 37 | | | | | 40104.188 | 39215.316 | 38077.477 |
| 38 | | | | | 40248.383 | 39487.903 | 38425.344 |
| 39 | | | | | 40344.513 | 39720.327 | 38738.534 |
| 40 | | | | | 40392.578 | 39912.587 | 39017.048 |
| 41 | | | | | | 40064.682 | 39260.884 |
| 42 | | | | | | 40176.615 | 39470.043 |
| 43 | | | | | | 40248.383 | 39644.526 |
| 44 | | | | | | 40279.987 | 39784.331 |
| 45 | | | | | | | 39889.459 |
| 46 | | | | | | | 39959.911 |
| 47 | | | | | | | 39995.685 |
| 48 | | | | | | | 39996.782 |

Table S5. Calculated vibrational energies ($G(\upsilon)$) for $^1H_2$–$^7H_2$ isotopologues in the M2($r$) potential.

| Vib. state ($\upsilon$) | $G(\upsilon)$, cm$^{-1}$ | | | | | | |
|---|---|---|---|---|---|---|---|
| | $^1H_2$ | $^2H_2$ | $^3H_2$ | $^4H_2$ | $^5H_2$ | $^6H_2$ | $^7H_2$ |
| 0 | 2168.415 | 1541.450 | 1262.482 | 1093.946 | 978.993 | 894.126 | 828.269 |
| 1 | 6315.652 | 4529.352 | 3723.960 | 3234.278 | 2898.948 | 2650.690 | 2457.638 |
| 2 | 10210.097 | 7390.594 | 6100.792 | 5311.199 | 4768.195 | 4365.005 | 4050.785 |
| 3 | 13851.753 | 10125.173 | 8392.977 | 7324.707 | 6586.734 | 6037.070 | 5607.708 |
| 4 | 17240.617 | 12733.091 | 10600.516 | 9274.803 | 8354.565 | 7666.886 | 7128.407 |
| 5 | 20376.691 | 15214.348 | 12723.408 | 11161.486 | 10071.689 | 9254.452 | 8612.884 |
| 6 | 23259.974 | 17568.942 | 14761.653 | 12984.758 | 11738.105 | 10799.768 | 10061.137 |
| 7 | 25890.466 | 19796.875 | 16715.251 | 14744.617 | 13353.813 | 12302.835 | 11473.167 |
| 8 | 28268.167 | 21898.147 | 18584.203 | 16441.064 | 14918.813 | 13763.652 | 12848.973 |
| 9 | 30393.078 | 23872.757 | 20368.508 | 18074.099 | 16433.105 | 15182.219 | 14188.556 |
| 10 | 32265.198 | 25720.705 | 22068.166 | 19643.721 | 17896.689 | 16558.537 | 15491.916 |
| 11 | 33884.527 | 27441.991 | 23683.178 | 21149.931 | 19309.566 | 17892.605 | 16759.053 |
| 12 | 35251.065 | 29036.616 | 25213.543 | 22592.729 | 20671.735 | 19184.423 | 17989.966 |
| 13 | 36364.813 | 30504.580 | 26659.262 | 23972.115 | 21983.196 | 20433.992 | 19184.656 |
| 14 | 37225.770 | 31845.881 | 28020.333 | 25288.089 | 23243.949 | 21641.311 | 20343.122 |
| 15 | 37833.936 | 33060.521 | 29296.758 | 26540.650 | 24453.994 | 22806.381 | 21465.366 |
| 16 | 38189.311 | 34148.500 | 30488.537 | 27729.799 | 25613.332 | 23929.201 | 22551.386 |
| 17 | 38291.896 | 35109.817 | 31595.668 | 28855.536 | 26721.961 | 25009.771 | 23601.182 |
| 18 | | 35944.472 | 32618.153 | 29917.861 | 27779.883 | 26048.091 | 24614.756 |
| 19 | | 36652.465 | 33555.991 | 30916.773 | 28787.097 | 27044.162 | 25592.106 |
| 20 | | 37233.797 | 34409.183 | 31852.274 | 29743.603 | 27997.984 | 26533.232 |
| 21 | | 37688.467 | 35177.728 | 32724.362 | 30649.401 | 28909.556 | 27438.136 |
| 22 | | 38016.476 | 35861.626 | 33533.037 | 31504.492 | 29778.878 | 28306.816 |
| 23 | | 38217.823 | 36460.878 | 34278.301 | 32308.875 | 30605.950 | 29139.272 |
| 24 | | 38292.508 | 36975.483 | 34960.152 | 33062.549 | 31390.773 | 29935.506 |
| 25 | | | 37405.441 | 35578.591 | 33765.516 | 32133.346 | 30695.516 |
| 26 | | | 37750.752 | 36133.618 | 34417.776 | 32833.670 | 31419.303 |
| 27 | | | 38011.417 | 36625.233 | 35019.327 | 33491.744 | 32106.866 |
| 28 | | | 38187.435 | 37053.435 | 35570.170 | 34107.568 | 32758.207 |
| 29 | | | 38278.807 | 37418.225 | 36070.306 | 34681.142 | 33373.323 |
| 30 | | | 38285.532 | 37719.603 | 36519.734 | 35212.467 | 33952.217 |
| 31 | | | | 37957.569 | 36918.454 | 35701.543 | 34494.887 |
| 32 | | | | 38132.123 | 37266.466 | 36148.369 | 35001.334 |
| 33 | | | | 38243.264 | 37563.770 | 36552.945 | 35471.558 |
| 34 | | | | 38290.993 | 37810.367 | 36915.271 | 35905.558 |
| 35 | | | | | 38006.256 | 37235.348 | 36303.335 |
| 36 | | | | | 38151.437 | 37513.175 | 36664.889 |
| 37 | | | | | 38245.910 | 37748.753 | 36990.219 |
| 38 | | | | | 38289.675 | 37942.080 | 37279.326 |
| 39 | | | | | | 38093.159 | 37532.210 |
| 40 | | | | | | 38201.987 | 37748.870 |
| 41 | | | | | | 38268.566 | 37929.307 |
| 42 | | | | | | 38292.896 | 38073.521 |
| 43 | | | | | | | 38181.512 |
| 44 | | | | | | | 38253.279 |
| 45 | | | | | | | 38288.823 |

Figure S2. The difference between the half of the energy of the fundamental transitions 0 → 1 and zero-point energy for $^{n}H_2$ ground state X $^{1}\Sigma_g^{+}$ as a function of *n*.

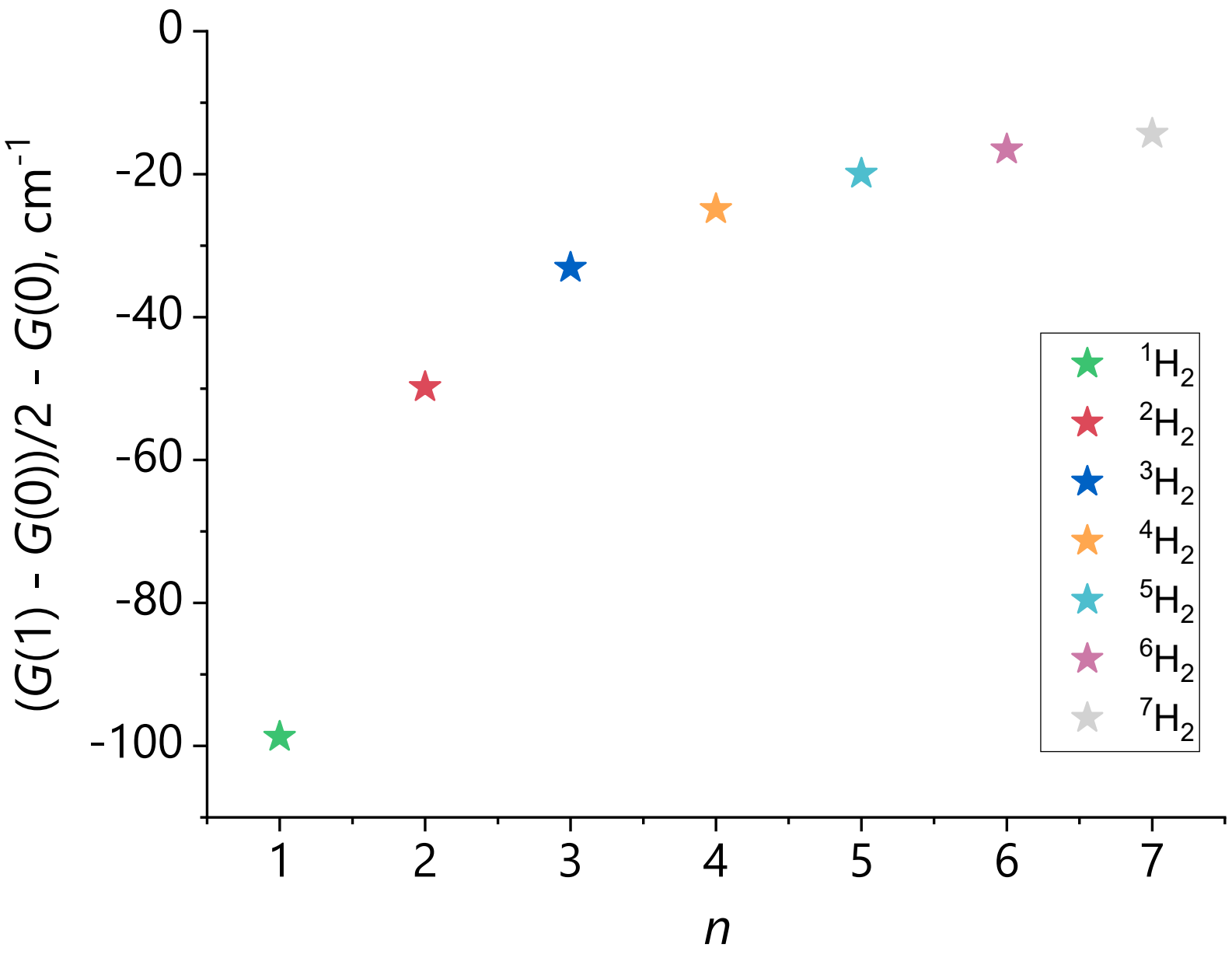

Table S6. Intersection points of the U(*r*) potential with its M1(*r*) and M2(*r*) approximations for $^1H_2$–$^7H_2$.

| | U(*r*)∩M1(*r*), Å | U(*r*)∩M2(*r*), Å |
|---|---|---|
| $^1H_2$ | 0.73299; 0.74063; 0.75767 | 0.73244; 0.74078; 0.76101; 1.25236 |
| $^2H_2$ | 0.73154; 0.74087; 0.75485 | 0.73023; 0.74094; 0.75692; 1.26031 |
| $^3H_2$ | 0.72997; 0.74099; 0.75391; 1.66332; 2.11923 | 0.72804; 0.74100; 0.75503; 1.26503 |
| $^4H_2$ | 0.73074; 0.74093; 0.75442; 1.62745; 2.17225 | 0.72797; 0.74102; 0.75483; 1.26536 |
| $^5H_2$ | 0.72959; 0.74101; 0.75379; 1.63892; 2.15805 | 0.72794; 0.74102; 0.75480; 1.26546 |
| $^6H_2$ | 0.72858; 0.74105; 0.75346; 1.60146; 2.22338 | 0.72753; 0.74106; 0.75434; 1.26702 |
| $^7H_2$ | 0.72732; 0.74108; 0.75275; 1.52922; 2.36976 | 0.72699; 0.74111; 0.75390; 1.26876 |